\documentclass[a4paper,11pt]{article}
\pdfoutput=1 % if your are submitting a pdflatex (i.e. if you have
\usepackage{jcappub} % for details on the use of the package, please
\usepackage{qubic}
\usepackage{tcolorbox}

\title{QUBIC IV:  Performance of TES Bolometers and Readout Electronics}
\author[1]{M.~Piat}
\author[1]{G.~Stankowiak}
\author[2,3]{E.S.~Battistelli}
\author[2,3]{P.~de~Bernardis}
\author[2,3]{G.~D'Alessandro}
\author[2,3]{M.~De~Petris}
\author[1]{L.~Grandsire}
\author[1]{J.-Ch.~Hamilton}
\author[4]{T.D.~Hoang}
\author[5]{S.~Marnieros}
% ...

\affiliation[1]{Universit\'e de Paris, CNRS, Astroparticule et Cosmologie, F-75013 Paris, France}
\affiliation[2]{Universit\a`{a} di Roma - La Sapienza, Roma, Italy}
\affiliation[3]{INFN sezione di Roma, 00185 Roma, Italy}
\affiliation[4]{University of Science and Technology of Hanoi, Vietnam Academy of Science and Technology}
\affiliation[5]{Laboratoire de Physique des 2 Infinis Ir\a`{e}ne Joliot-Curie (CNRS-IN2P3, Universit\a'e Paris-Saclay), France}
% ...

\emailAdd{piat@apc.univ-paris7.fr}
\emailAdd{stankowi@apc.in2p3.fr}

\abstract{
{A prototype version of the Q \& U {bolometric interferometer for cosmology} (QUBIC) underwent a campaign of testing in the laboratory at
Astroparticle Physics and Cosmology {laboratory} in Paris {(APC)}. The detection chain is currently made of 256 NbSi {transition edge sensors (TES)} cooled to 320~mK. The readout system is a 128:1 time domain multiplexing scheme based on 128 SQUIDs cooled at 1~K that are controlled and amplified by an SiGe {application specific integrated circuit} at 40~K.  We report the performance of this readout chain and the characterization of the TES. The readout system has been functionally tested and characterized in the lab and in QUBIC. The {low noise amplifier} demonstrated a white noise level of 0.3~$\mathrm{nV}/\sqrt{\mathrm{Hz}}$. Characterizations of the QUBIC detectors and readout electronics includes the measurement of I-V curves, time constant and the {noise equivalent power}. The QUBIC TES bolometer array has approximately 80\% detectors within operational parameters. {It demonstrated a thermal decoupling compatible with a phonon noise of about $5\times10^{-17}~\mathrm{W}/\sqrt{\mathrm{Hz}}$ at 410~mK critical temperature.}  While still limited by microphonics from the pulse tubes and noise aliasing from readout system, the {instrument noise equivalent power} is about
$2\times10^{-16}~\mathrm{W}/\sqrt{\mathrm{Hz}}$, enough for the demonstration of bolometric interferometry.
}

}

\date{\today}

\begin{document}

\maketitle

\section{Introduction}
\label{sec:intro}

QUBIC is an international ground based experiment dedicated to the observation of {cosmic microwave background} (CMB) polarisation. It will be deployed in Argentina, at the Alto Chorrillos mountain site (altitude of 4869~m a.s.l.) near San Antonio de los Cobres, in the Salta province. 
QUBIC has two configurations: the {``technological demonstrator''} (TD)  and the {``full instrument''} (FI). The TD and FI share the same cryostat and cryogenics but the TD has only one-quarter of the 150~GHz TES focal plane (256 TESs), an array of 64 horns and switches and a smaller optical combiner. The QUBIC TD has demonstrated the feasibility of the bolometric interferometry after extensive tests at APC laboratory since 2018. In this paper, we present the main results of this characterization phase on the detection chain. 
%First section will describe briefly the QUBIC detection chain.

%The QUBIC detection chain architecture is described in detail in \cite{2016arXiv160904372A}. The QUBIC-TD focal plane is based on one 256-pixel array operating at 150~GHz (a quarter of QUBIC focal plane). Two blocks of 128~SQUIDs are used at 1~K in a 128:1 Time Domain
%Multiplexing scheme. Each block is controlled and amplified by an ASIC cooled to 40~K while a warm FPGA board takes care of the control and acquisition of the signal to the acquisition computer.

%To check the performance of the QUBIC-TD detection chain, we characterized the system in the normal state, in the superconducting state, and in the transition state. Unless otherwise specified, the measurements shown here were performed with a temperature regulation on the TES stage at 362~mK.

This paper is organized as follows.  An overview of the QUBIC detection chain is given in Section~\ref{sec:detchain}. The tests of the readout system is described in Section~\ref{sec:readout_charact}.
Section~\ref{sec:TEScharacterization_functionality} describes the TES characterizations in terms of critical temperature, TES parameters, power background, time constants and noise performance. 
%Section~\ref{sec:TEStimeconstant} describes the
%measurement of the TES time constant using the external modulated
%calibration source.  In Section~\ref{sec:linearity} the linearity of
%the TES is evaluated.  Section~\ref{sec:magfield} discusses
%measurements of magnetic field pickup by the SQUIDs and the mitigation
%solution which was implemented.  
Finally, some concluding remarks are
given in Section~\ref{sec:conclusion}.

%\clearpage
\section{QUBIC detection chain}
\label{sec:detchain}

\begin{figure}[htbp]
\begin{center}
{\includegraphics[width=0.6\linewidth, keepaspectratio]{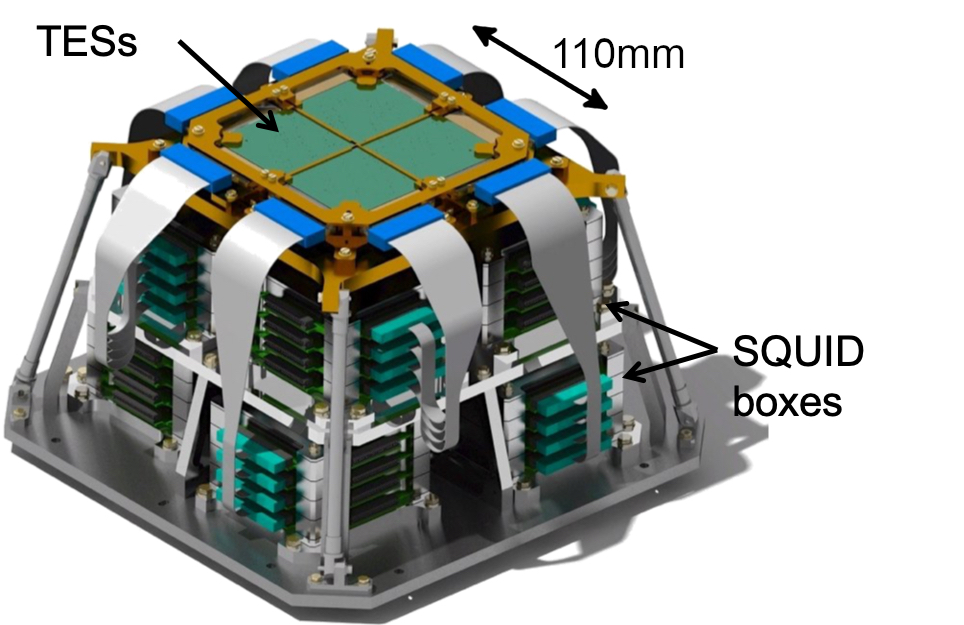}}
{\includegraphics[width=0.6\linewidth, keepaspectratio]{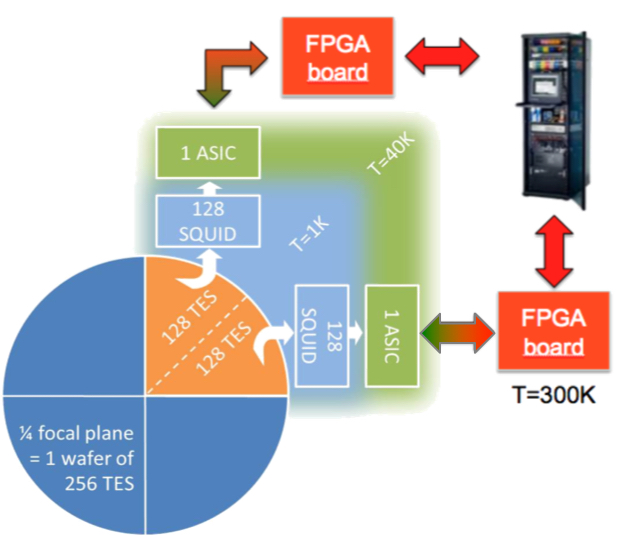}}
\caption{{\it top}: QUBIC cryo-mechanical structure which supports one TES focal plane at 350~mK on top and the SQUID boxes at 1~K below.  {The focal plane diameter is 110~mm. }{\it bottom}: Architecture of the QUBIC detection chain for one focal plane of 1024 channels, highlighted on one quarter of it.
\label{Archi_det_chain}}
\end{center}
\end{figure}

The QUBIC detection chain architecture is shown on Figure~\ref{Archi_det_chain}. % and is described in detail in \cite{2016arXiv160904372A}. 
Each focal plane is composed of four 256-pixel TES arrays assembled together to obtain 1024-pixel detector cooled at about 320~mK by a $^3$He fridge. For each quarter focal plane, two blocks of 128 SQUIDs ({superconducting quantum interference devices}) are used at 1~K in a 128:1 {time domain multiplexing} (TDM) scheme  \cite{2018JLTP..193..455P, 2020JLTP..200..363P}. Each block is controlled and amplified by an ASIC ({application specific integrated circuit}) cooled to 40~K while a warm FPGA ({field programmable gate array}) board ensure the control and acquisition of the signal to the acquisition computer. 

\subsection{TES}
\begin{figure}[htbp]
\begin{center}
{\includegraphics[scale=0.55]{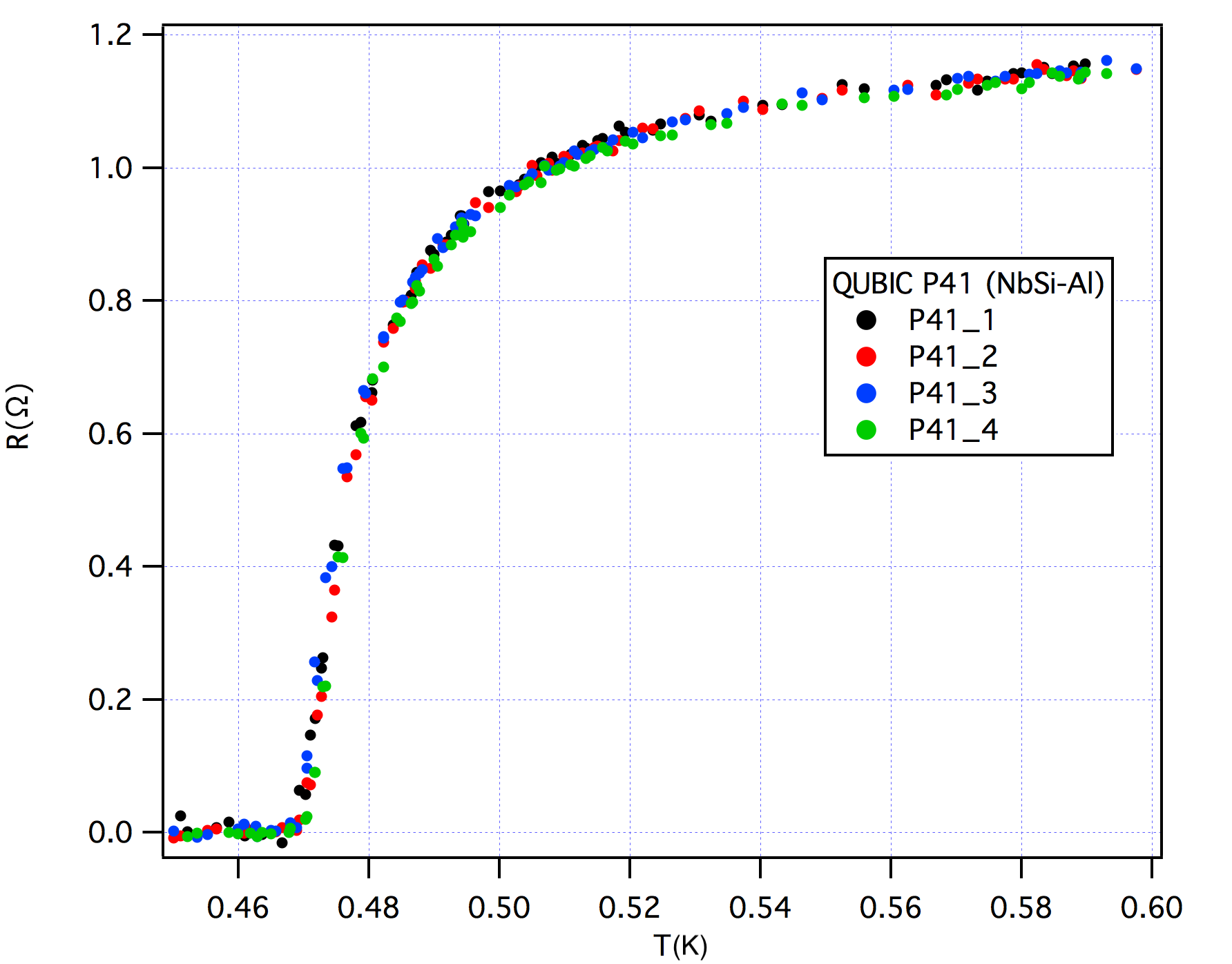}}
\caption{Superconducting transition characteristics (resistance R versus temperature T) of four Nb\textsubscript{0.15}Si\textsubscript{0.85} TESs distributed far away from each other on the 256 pixel array reference P41.
\label{testransi1}}
\end{center}
\end{figure}

The detectors are {TESs} made with a Nb\textsubscript{x}Si\textsubscript{1-x} 
amorphous thin film (x${\approx}$0.15 in our case). 
Their transition temperature T\textsubscript{c} %of about 500 mK$
(Figure~\ref{testransi1}) can be adapted by changing the composition $x$ of the compound. The array currently used (reference P87) has a critical temperature of about 410~mK. The normal state resistance R\textsubscript{n} is adjusted to about 1~$\Omega$ with interleaved electrodes for optimum performance. To adapt to the optics, the pixels have 3~mm spacing while the grid absorber structure is 2.7~mm wide without sensitivity to polarization.
%\includegraphics[height=4.3cm, keepaspectratio]{figures/R(T)}
%The critical normal-to-superconducting temperature is close to 500 mK, as illustrated by Figure~\ref{testransi}. Voltage biasing of the sensors allows operation on the well known ``extreme electro-thermal feedback'' mode with increased bandwidth, direct power calibration and self-regulation of the TES at the superconducting transition temperature. 
%
%Given the expected  background power of the QUBIC setup (5-50 pW in the 150-220 GHz range) an
%extremely low thermal coupling between the TES and the cryostat is needed to optimize signal to noise ratio. 
The low thermal coupling between the TES and the thermal bath is obtained using 500~nm thin SiN suspended and patterned membranes, which exhibit thermal {conductance} in the range 50-500~pW/K depending on the precise pixel geometry and temperature. The {noise equivalent power} (NEP) is expected to be of the order of $5\times10^{-17}~W/\sqrt{Hz}$ at 150~GHz with a natural time constant of about 100~ms \cite{2018SPIE10708E..45S}. 
Light absorption is achieved using a Palladium metallic grid placed in a quarter-wave
cavity optimizing the  absorption efficiency. The back-short distance of 400 ${\mu}$m has been chosen after electromagnetic simulations in order to have absorption higher than 94\% at both 150 and 220~GHz. 
%The array is not intrinsically sensitive to polarization.
The routing of the signal between the TES and the bonding pads at the edge of the array is realised by superconducting aluminium lines patterned on the silicon frame supporting the membranes. The detailed fabrication process of the QUBIC detectors is given in  \cite{2020JLTP..199..955M}. The latest upgrade of the production process allows excellent fabrication quality with a dead-pixel yield as low as 5\% at room temperature. 
\begin{figure}[htbp]
\begin{center}
{
\begin{minipage}[t]{\linewidth}
\includegraphics[height=6.3cm, keepaspectratio]{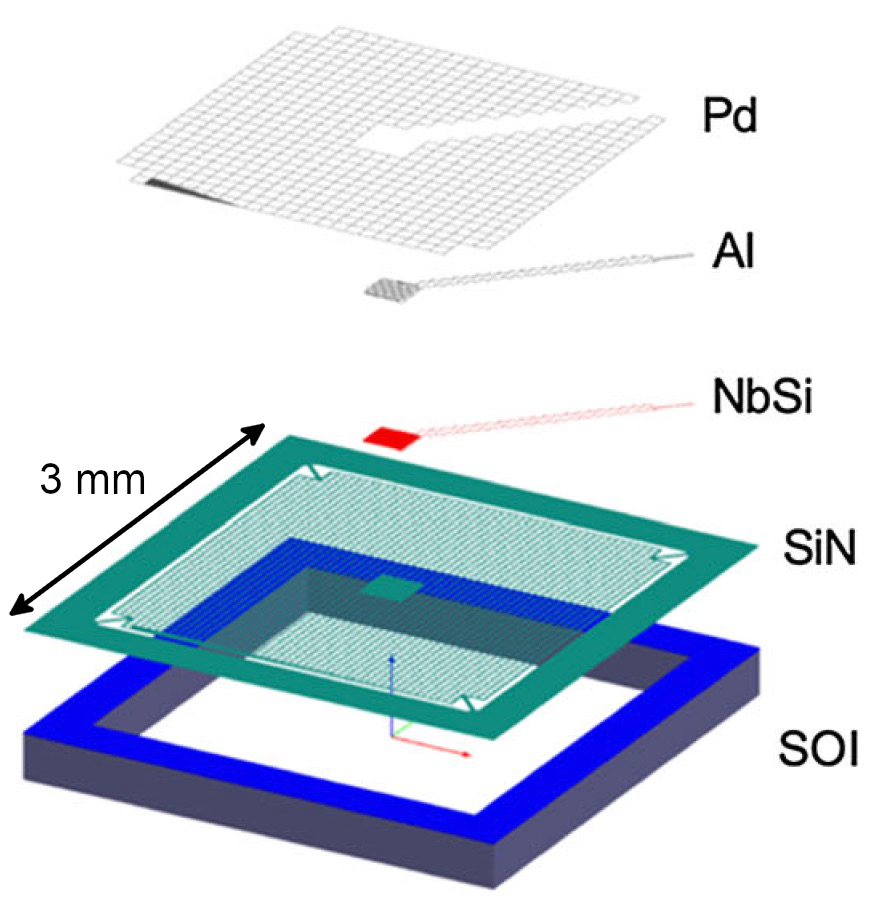}
\includegraphics[height=6.3cm, keepaspectratio]{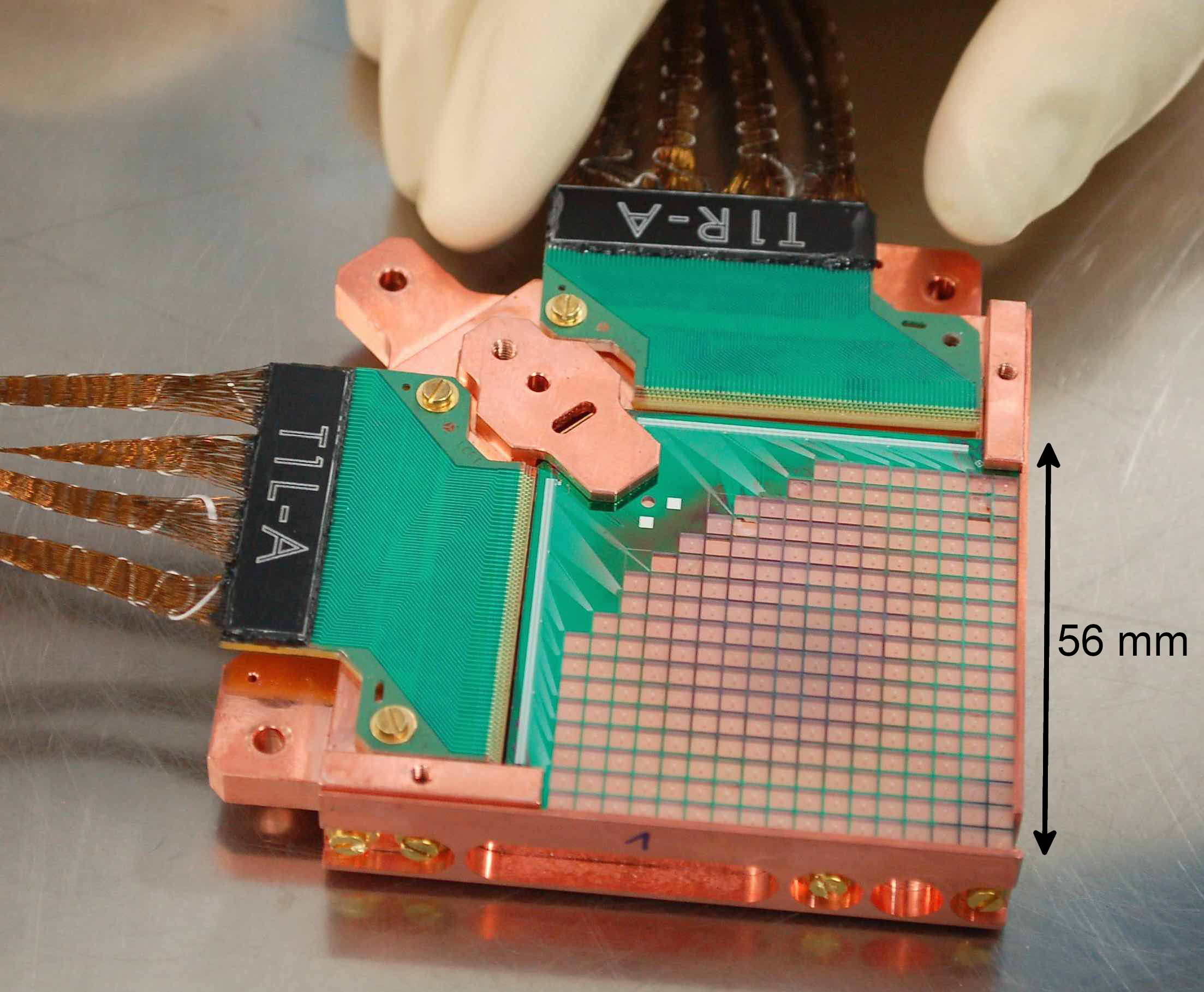} 
\end{minipage}}
%{\includegraphics[height=5.3cm, keepaspectratio]{figures/TESgrid_new.png}} 
%\includegraphics[height=4.3cm, keepaspectratio]{figures/QUBICTDRcompilation-img140.jpg} 
\caption{
{\it Left}: Layout of the 3-mm pitch pixel structure. Pd
grid for light absorption, NbSi for temperature sensing, SiN structure for decoupling the sensor from the
cold bath and Al for routing the signal to the SQUID amplifiers
{\it Right}: A 256 TES array  being integrated.
\label{testransi}}
\end{center}
\end{figure}
\\
The 256-pixel array is finally integrated within the focal plane holder and electrically connected to an aluminium {printed circuit board} (PCB, provided by Omni Circuit Boards\footnote{www.omnicircuitboards.com}) using ultrasonic bonding of aluminium wires (Figure~\ref{testransi}). 

\subsection{SQUIDs}
\begin{figure}[ht!]
	\centering
    	\includegraphics[width=0.7\linewidth]{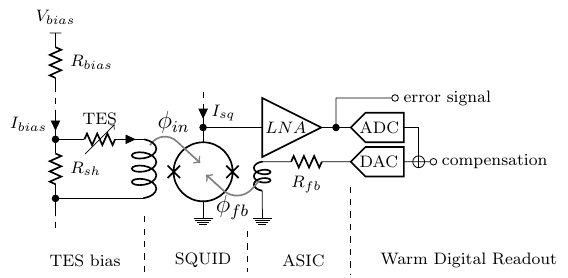}
    \caption{\label{fig:squidFLL} Layout of the TES, SQUID and ASIC operating in flux-locked loop.}
\end{figure}

The {second stage of the} detection chain is composed of the {SQUIDs} maintained at a temperature of about 1~K by a $^4$He fridge. Each TES is in series with the input inductance $L_\mathrm{in}$ of the SQUID and is voltage biased with a $10~m\Omega$ resistor in parallel as shown in Figure~\ref{fig:squidFLL}. The input inductance of the SQUID converts the TES current into a magnetic flux $\Phi_\mathrm{in}$ that is converted in an output voltage by the SQUID. The latter is therefore  a trans-impedance amplifier with a gain of the order of 100~V/A. 
To compensate the flux variation, a current from a feedback loop is injected to create a feedback flux $\Phi_\mathrm{fb}$. The voltage sent by the {digital to analog converter (DAC)} to create this feedback current through the feedback resistor $R_\mathrm{fb}$ is the QUBIC signal (Figure~\ref{fig:squidFLL}). This process, allowing to lock the flux operating point in the SQUID is known as a {flux locked loop (FLL)} \cite{FLL}. 

\begin{figure}[ht!]
    \centering
    {
    \begin{minipage}[t]{\linewidth}
    \includegraphics[width=0.6\linewidth]{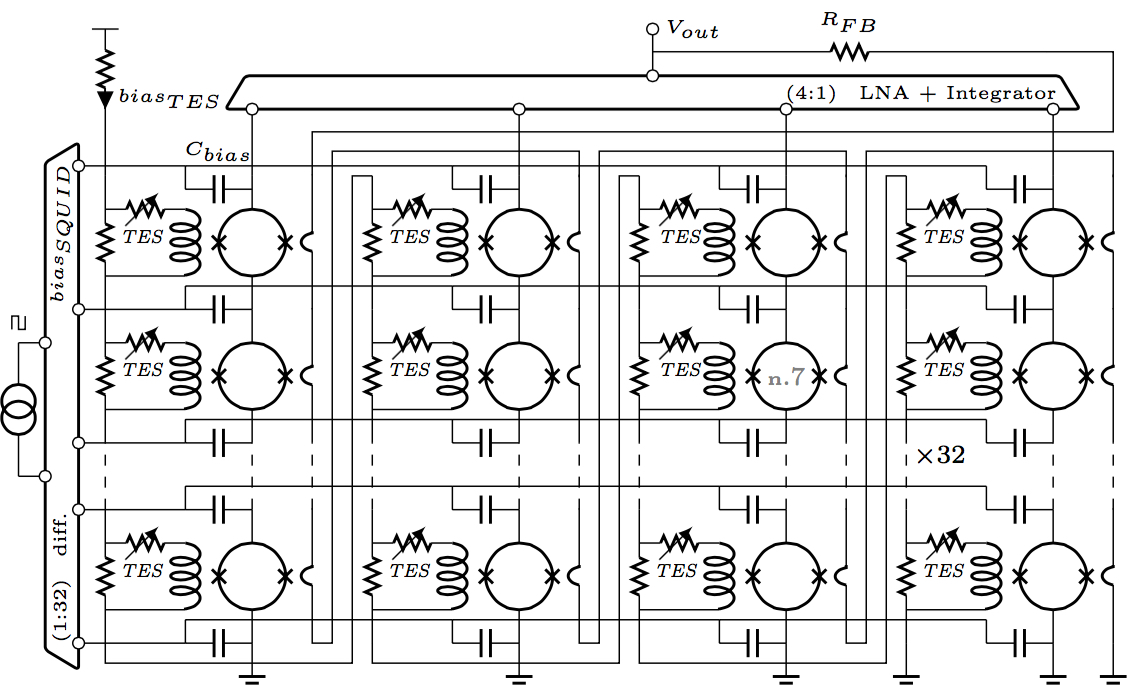}
    \includegraphics[width=0.35\linewidth]{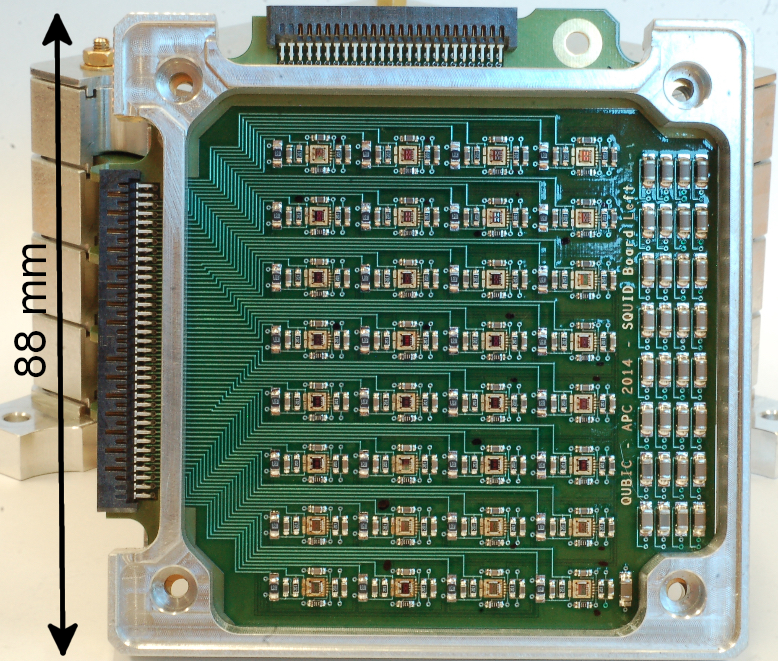}
    \end{minipage}}
    \caption{\label{fig:detection_chain_readout}{\it Left}: Topology of the 128 to 1 multiplexer sub-system (4$\times$32 SQUIDs + 1 ASIC). {\it Right}:~~Integration of 32 SQUIDs (1 column) with bias capacitors and filter devices.}
\end{figure}

\begin{figure}[ht!]
	\centering
    	{\includegraphics[width=0.4\linewidth, angle =90, trim={0 5cm 0 0}, clip ]{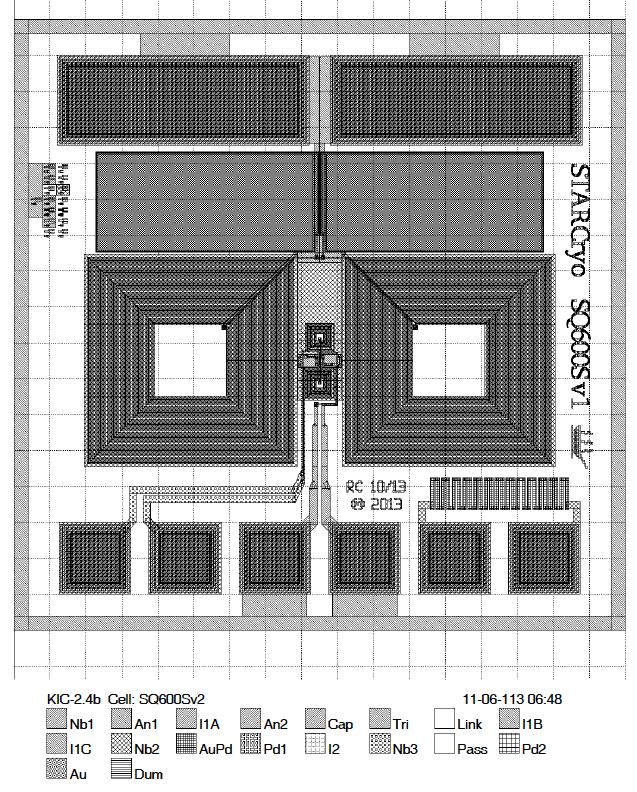}
    	\includegraphics[width=0.4\linewidth ]{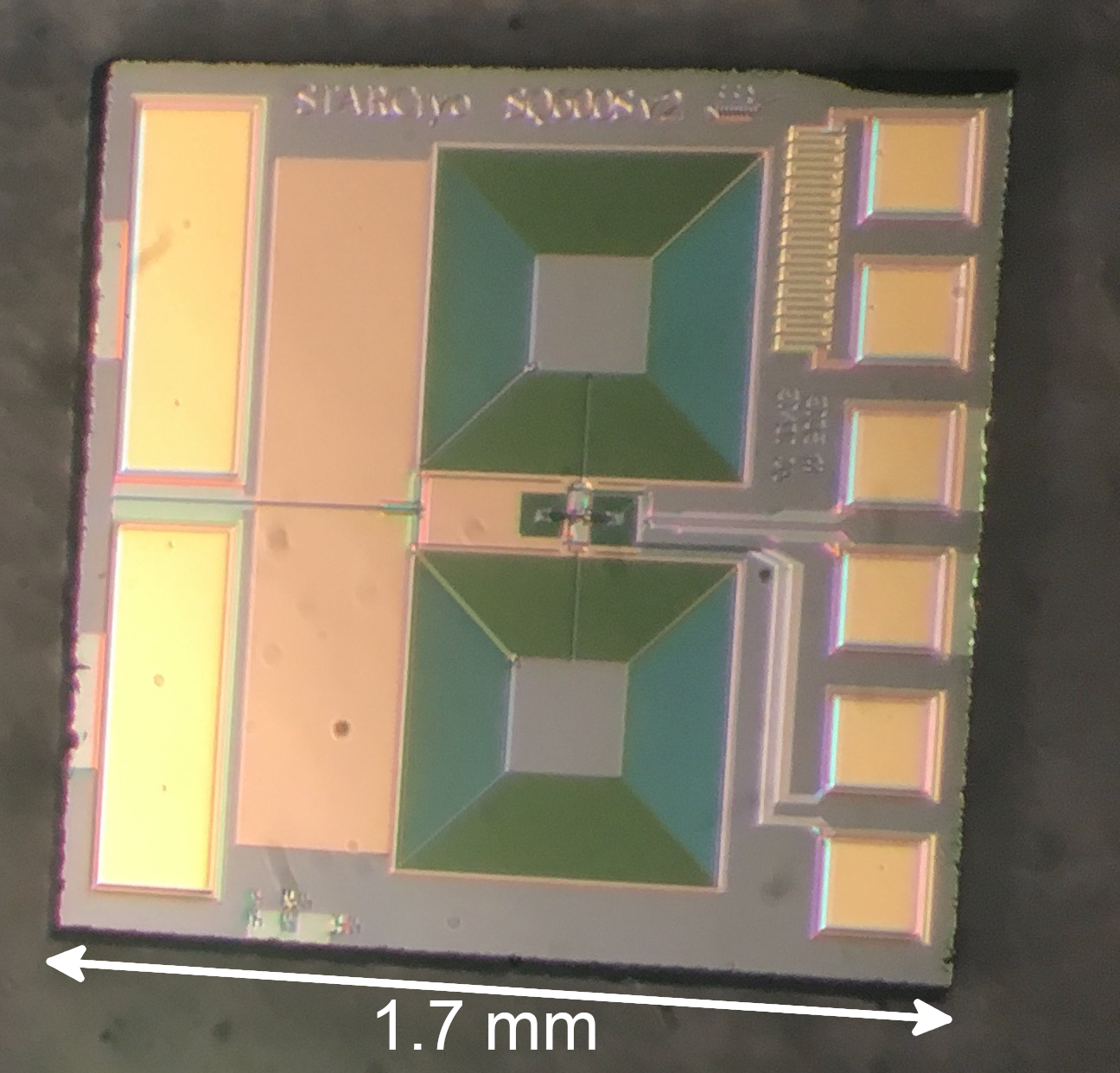}}
    \caption{\label{fig:squidphoto}On the left, layout of a gradiometric SQUID (100~$\mu m$ grid), on the right, picture of one SQUID bare die (about 1.7 mm side).}
\end{figure}

In addition to being cryogenic amplifiers, SQUIDs also enable the multiplexing because of their large bandwidth. 
As shown in Figure~\ref{fig:detection_chain_readout}, the SQUID multiplexer is composed of 4~columns of 32~SQUIDs AC-biased with capacitors in order to reduce power dissipation and noise.
The SQUIDs used in QUBIC shown in Figure \ref{fig:squidphoto} have a dual-washer gradiometric layout. They are based on a SQ600S commercial design provided by StarCryoelectronics\footnote{starcryo.com}, slightly modified in order to reduce the area of each die.

%However, this design has been modified to remove an input transformer (for ``current-lock'' CL operation) not used in the QUBIC readout chain (based on flux feedback). In addition, size of the pads has been reduced to $200\mu$m and all the design has finally be compacted to reduce the area for each SQUID and put about 4000 SQUIDs on 2 custom wafers.
%
Visual inspections and room temperature tests with a probe-station are used to select the SQUIDs before integration on a specific PCB. One SQUID PCB is composed of 32~SQUIDs and is integrated in an aluminium box. The architecture therefore uses 4 of these PCB boxes to read out 128 pixels. As shown in Figure \ref{Archi_det_chain}, 4 stack of 8 SQUID boxes is installed at 1~K below the TESs in the cryo-mechanical structure, surrounded with a Cryophy\footnote{www.aperam.com} magnetic shield.

% The QUBIC detection chain second stage is maintained at a temperature of 1~K and is composed of the Superconducting QUantum Interference Devices (SQUID).  The TES voltage biasing leads to a current readout. The voltage is kept constant across the TES.  We measure the fluctuation of the current induced by the TES resistance fluctuation. The input impedance of the SQUID must be smaller than shunt resistance, $R_\mathrm{shunt}$, in order to avoid addition of noise in the biasing circuit.  The input impedance of the SQUID corresponds to the impedance of the SQUID input loop $L_\mathrm{in}$. The input loop is made using superconducting material (Nb) which does not add any resistance to the TES biasing circuit.  The impedance of the input loop, $L_\mathrm{in}$, converts the TES current in flux, $\Phi_\mathrm{in}$, into the SQUID which, in turn, provides an output voltage with trans-impedance amplification (gain) of the order of 100~V/A.  This gain is strongly non-linear as seen in Figure~\ref{fig:V_Phi_response}.  A feedback flux is applied through the feedback coil and the feedback resistance in order to counteract the input flux from TES current fluctuation.  This feedback technique provides a wide linear range to readout the TES.
%\begin{figure}[ht!]
%    \centering
%    \includegraphics[width=1\linewidth]{V(Phi)_QUBIC}
%    \caption{\label{fig:V_Phi_response}SQUID non linear response}
%\end{figure}

\subsection{{ASIC}}

In addition to the SQUIDs, a 4~to~1 multiplexed low noise amplifier (LNA) reads out sequentially 4~columns of 32~SQUID each.  The resulting multiplexing factor is~128.  The LNA, together with sequential biasing of the SQUID and the overall clocking of this 128:1 multiplexer, is all done in a cryogenic ASIC operating at about 40~K \citep{2016JLTP..184..363P}.  The TDM readout is based on the association of 4~columns of 32~SQUIDs in series with the dedicated cryogenic ASIC.

The ASIC is designed in full-custom using CADENCE CAD tools. 
The technology is a standard $0.35~\mu$m  BiCMOS SiGe from Austria MicroSystem (AMS).  
This technology consists of p-substrate, 4-metal and 3.3~V process. It includes standard complementary MOS transistors
and high speed vertical SiGe NPN {hetero-junction bipolar transistors} (HBT). Bipolar transistors are preferentially used
for the design of analog parts because of their good performance at cryogenic temperature \cite{2018JLTP..193..455P}. 
The design of the ASIC is based on pre-experimental characterizations results, and its performance at cryogenic temperature is extrapolated from simulation results obtained at room temperature, using CAD tools.
\begin{figure}[htbp]
\begin{center}
{
\raisebox{0.25\height}{\includegraphics[width=0.45\linewidth, keepaspectratio]{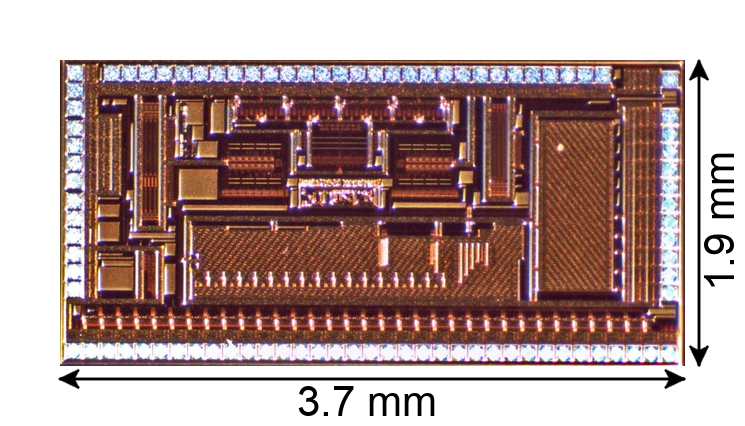}}
\includegraphics[width=0.45\linewidth, keepaspectratio]{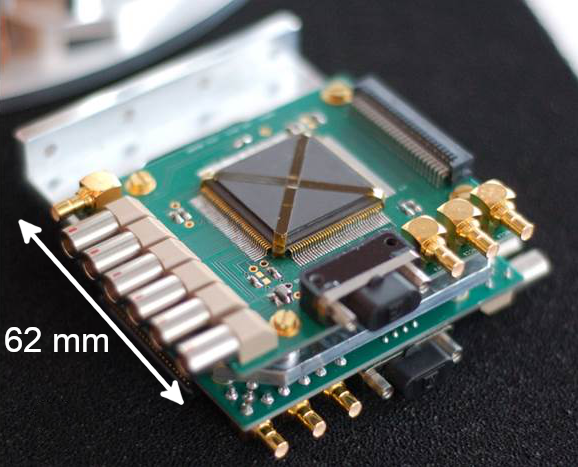}}
\caption{{\it Left}:  Microphotography of the cryogenic ASIC
designed to read out 4$\times$32 TES/SQUID pixels. {\it Right}:~~ASIC module assembly used for QUBIC.
\label{fig:cryoasic}}
\end{center}
\end{figure}

Each ASIC board (shown in Figure~\ref{fig:cryoasic}) has a power dissipation of typically 16~mW and is placed on the 40-K stage. The ASIC integrates all parts needed to achieve the readout, the multiplexing and the control of an array of up to 128 TESs/SQUIDs. 
%It operates from room temperature down to 4.2K, thanks to a low power dissipation (16 mW per ASIC typically, whatever the number of columns to readout). 
It includes a differential switching current source to address sequentially 32 lines of SQUIDs,
achieving a first level of multiplexing of 32:1. In this configuration, the SQUIDs are AC biased through capacitors which allows good isolation (low cross-talk between SQUID columns) and no power dissipation. 
A cryogenic SiGe low-noise amplifier ($e_n=$~0.3~nV/$\sqrt{Hz}$, gain~=~70, bandwidth of about 6~MHz in simulations) with 4 multiplexed inputs, performs a second multiplexing stage between each {of the 4 columns}. The low frequency noise of the LNA increases with decreasing temperature. An operation at about 40~K appears to be a good trade-off between this corner frequency and the white noise level. 

This cryogenic ASIC includes also the digital synchronization circuit of the overall multiplexing switching (AC current sources and multiplexed low-noise amplifier). A serial protocol allows focusing on sub-array as well as adjusting the amplifiers and current sources with a reduced
number of control wires. 
%As the digital side takes a large part, 
We have developed a full-custom CMOS digital library dedicated to cryogenic applications and ionizing environments (rad-hard full custom digital library) \cite{2018JLTP..193..455P}.

%More than a cryogenic amplifier, SQUIDs also enable the multiplexing because of their large bandwidth. SQUID stages of 32~TES are connected together to readout successively each of the 32~TES. In addition, a 4~to~1 multiplexed LNA reads out sequentially 4~columns of 32~SQUID each.  The resulting multiplexing factor is~128.  The low noise amplifier (LNA) together with sequential biasing of the SQUID and the overall clocking of this 128:1 sub multiplexer is all done in a cryogenic Application Specific Integrated Circuit (ASIC) operating at 40~K \citep{2016JLTP..184..363P}.  The Time Domain Multiplexer (TDM) readout is based on the association of 4~columns of 32~SQUIDs in series with the dedicated cryogenic ASIC.

\subsection{{Warm electronics and acquisition software}}
The final stage of the readout electronics operates at room
temperature on a board called NetQuiC. It is connected to the acquisition computer via a network switch. 
Each NetQuiC board is based on a differential amplifier (gain~=~100, bandwidth limited to 1~MHz with a second-order anti aliasing low-pass filter), a 2~MHz 16-bit {analog to digital converter (ADC)}, seven 16-bit DACs and a Xilinx Spartan 6~FPGA (XEM6010 board from Opal Kelly). 
It also contains 2 feedback resistors $R_\mathrm{fb}$ of 10~k$\Omega$ and 100~k$\Omega$ that could be individually connected for large dynamic range or sensitive measurements respectively. 
This system is designed to adjust the operating biasing of the TESs and to control the feedback of the SQUIDs.  Moreover, it takes the signal from the cryogenic multiplexing ASIC, computes the scientific signal and sends it to the data acquisition system. For this detection chain each FPGA manages 128~detectors, with a total of
16~FPGAs for the full 2048~pixel focal planes.  
A dedicated software
named QUBIC~Studio was developed at the Institute for Research in
Astrophysics and Planetology (IRAP) for the data acquisition \cite{2016arXiv160904372A, 2020JLTP..200..363P}.
QUBIC~Studio interfaces with the generic {electrical ground support
equipment} (EGSE) tool, called {``dispatcher''}, which was also developed
at IRAP.  QUBIC~Studio has a user-friendly interface to manage the
connection with the readout electronics.  This tool gives a global
visualization of the complete detection chain.

%\clearpage
\section{Readout tests and characterization }
\label{sec:readout_charact}

The core of the readout is made of an ASIC cooled to 40~K that controls the multiplexing and amplifies the voltage from the SQUIDs. This device has been first tested and validated since it has been used to further characterize the SQUIDs.

\subsection{ASIC tests and characterizations}

\subsubsection{Implemented functions}
\begin{figure}[ht!]
    \begin{center}
    {\includegraphics[width=0.9\linewidth]{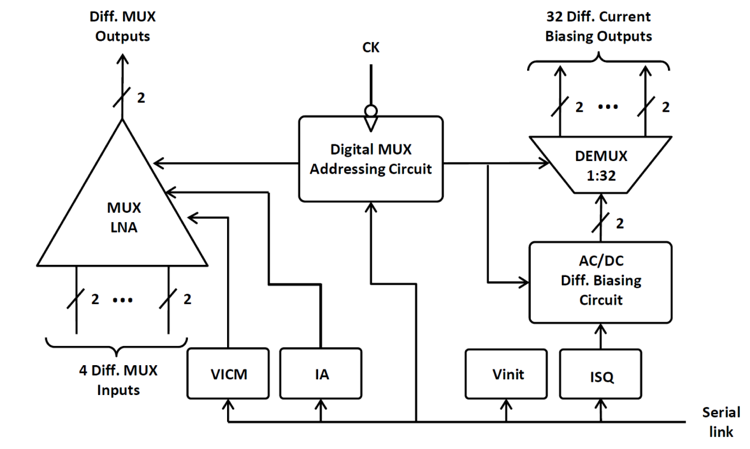}}
    \caption{Block diagram of the ASIC {\it SQMUX128evo} (see text for a detailed description).}
    \label{fig:ASIC_schem}
    \end{center}
\end{figure}

The ASIC {called} {\it SQMUX128evo} has been designed to control the time-domain multiplexing and to amplify the signals from 4~columns of 32~SQUIDs in series (see Figure \ref{fig:detection_chain_readout}) i.e. 128~channels. Its block diagram is outlined on Figure \ref{fig:ASIC_schem}. 
The following functions have been integrated:
\begin{itemize}
\item An ultra low noise voltage amplifier with 4~multiplexed inputs for reading 4~columns of SQUIDs,
\item a current source for the polarization of the multiplexed amplifier,
\item an AC bias current source associated with a 1:32 multiplexer for addressing the 32~SQUIDs lines through addressing capacitors,
\item two voltage references for adjusting the common mode at the input of the multiplexed amplifier and at the output of the AC bias source of the SQUID,
\item a digital circuit which controls the row / column addressing of the multiplexer from an external clock signal CK,
\item a serial link for addressing configurable blocks such as voltage references, bias current sources or the multiplexer's row/column addressing circuit.
\end{itemize}

%This ASIC has been made in standard BiCMOS SiGe 0.35~$\mu$m AMS technology and is powered at 3.3~V. 
{This ASIC} has been integrated on a specific PCB in order to be fully characterized at room and cryogenic temperatures. {It} has been tested and proven functional down to 4.2~K thanks to a low power dissipation of about 16~mW per ASIC whatever the number of columns to read out. In QUBIC, the ASIC operates at approximately 40~K due to the available cryogenic power.

\subsubsection{Current sources and voltage references}
The ASIC {\it SQMUX128evo} integrates digitally adjustable current sources and voltage references for the biasing of the multiplexed input amplifier and for the SQUID AC bias circuit.
The current sources are based on a fixed current reference ($I_{REF} \simeq 100~\mu$A typically) followed by current DACs whose values are encoded on 3 and 4 bits for the amplifier bias and the SQUID AC bias circuit respectively. The reference current $I_{REF}$ is obtained by taking the current flowing through an external resistor $R_{REF}$ under a fixed voltage (forward-biased diode voltage, about 0.7~V at room temperature). This allows to precisely adjust the reference current  value according to the operating temperature. For a nominal current $I_{REF} = 100~\mu $A:
\begin{itemize}
    \item The output of the 3-bit current DAC allows to adjust the bias of the amplifier with multiplexed inputs (IA in Figure~\ref{fig:ASIC_schem}) from 1.65~mA to 2.85~mA in steps of 200~$\mu$A;
    \item The output of the 4-bit DAC in current adjusts the AC bias of the SQUIDs (ISQ in Figure~\ref{fig:ASIC_schem}) from 5~$\mu $A to 40~$\mu $A in steps of 2.5~$\mu $A.% (referred to as indexes from 0 to 15 in the following).
\end{itemize}
The ASIC {\it SQMUX128evo} also incorporates two 3-bits voltage references for common mode adjustment at the input of the multiplexed amplifier (VICM) and at the output of the SQUID AC bias source (Vinit). This voltage ranges from 1.453~V to 1.895~V.

For the voltage references and current sources, the values measured at room temperature are fully compliant to those simulated. At low temperatures, an adjustment of the reference resistance and of the threshold voltage of a forward-biased diode from 0.7~V to about 1~V needs to be done to reproduce the results in simulation for the current sources.
The agreement between measurement and simulation {has been achieved thanks to the use of} a proven standard technology with a reliable design kit.

\subsubsection{Amplifier with 4 multiplexed inputs}
The amplifier is made of 4 differential pairs of SiGe bipolar transistors (each with a trans-conductance $g_m$) whose collectors are connected to a common resistor ($R_L~=~560~\Omega$ at room temperature and $500~\Omega$ at 40~K). The multiplexing is achieved by means of a set of CMOS switches that sequentially bias each differential pair that has to be activated ($I_{BIAS}~=~2~mA$ typically).
%The multiplexing of the inputs is done by switching on or off the bias current source ($I_{BIAS} = 2 mA$ typically) at the bottom of each differential pairs by means of a set of CMOS switches. 
A common mode (VICM) is applied at the input of each differential pair through 2 series resistors of $50~\Omega$ each (differential input impedance of $100~\Omega$) connected to a 3-bit voltage reference. Each output of the differential stage is followed by a common collector amplifier in order to reduce the output impedance to about $50~\Omega$ at low temperature. The expected maximum gain is about 20 and 70 at room and cryogenic temperature respectively.

%The contribution of the first two terms is minimized by using a large number of transistors in parallel to reduce the value of the intrinsic base resistance Rbi. Noise performance optimization is then reached by determining IC and RL to minimize the predominant contribution of the last two terms.
%The expression of the differential voltage gain is $A_{VD}\simeq -g_m \times R_L$

Gain and noise measurements were performed using an Agilent\footnote{http://www.agilent.com} HP89410 vector analyser. For the gain measurement, as the vector analyser does not have differential sources and inputs, the setup uses a "single to differential" circuit, made from an AD8132, to differentiate the signal coming from the analyser source and drive the input of the amplifier under test. The output common mode of the AD8132 is adjusted to match the VICM of the amplifier under test. A Stanford Research SR560 amplifier is used to differentiate between the outputs of the amplifier under test before feedback to the input of the analyser. This external amplifier limits the bandwidth to about 1~MHz. For noise measurement, this amplifier is also used to provide the extra gain needed to overcome the noise floor of the analyser. In addition, the noise measurement is performed by shunting the differential inputs of the amplifier under test with a short circuit in the lab or with zero bias of the SQUIDs in QUBIC.
The amplifier with multiplexed inputs is biased at maximum current (1.80~mA at 300~K and 2.85~mA at 77~K) so that the voltage gain is as high as possible. 

\begin{figure}[ht!]
    \begin{center}
{\includegraphics[width=0.7\linewidth]{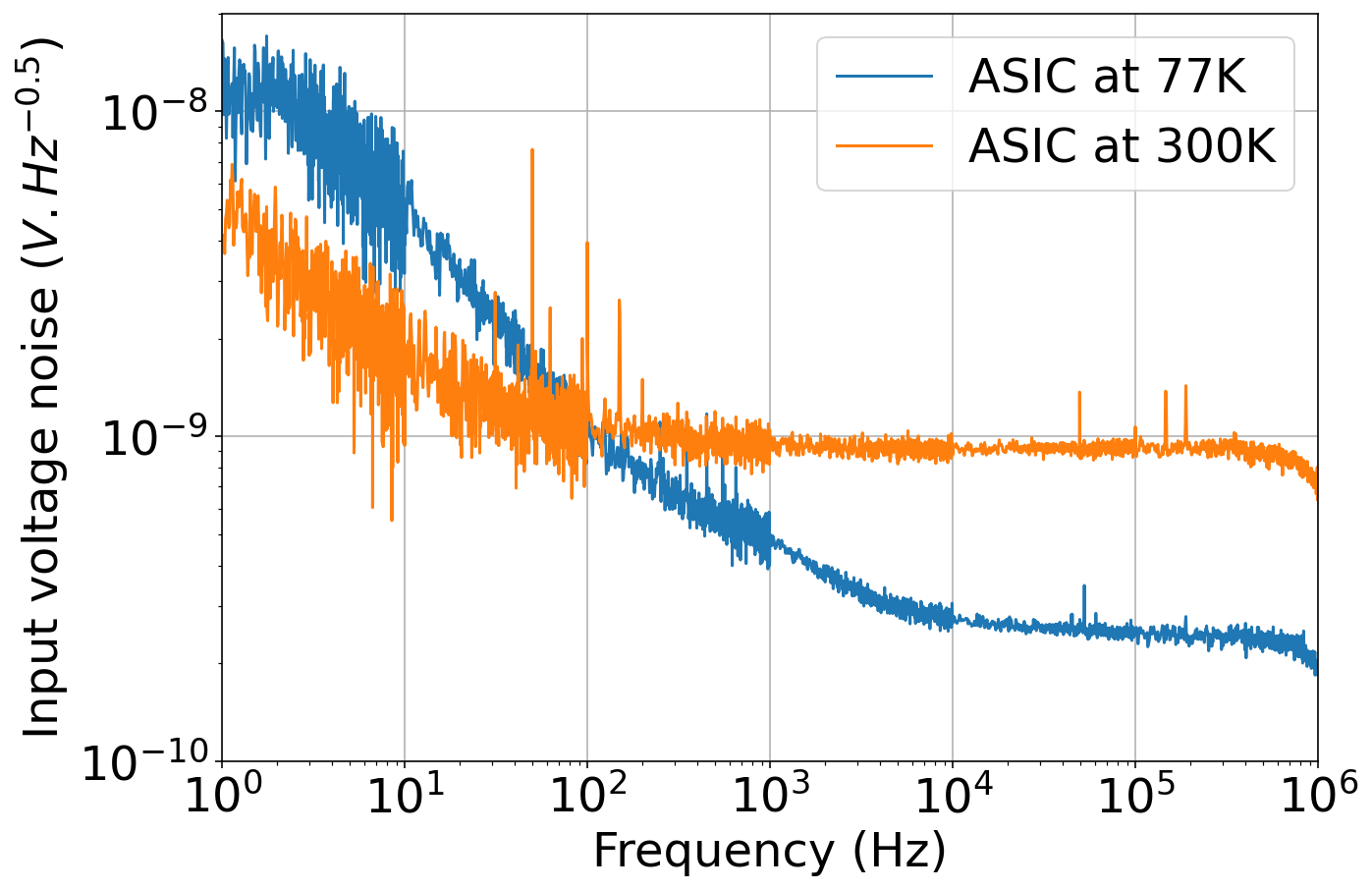}}
{\includegraphics[width=0.7\linewidth]{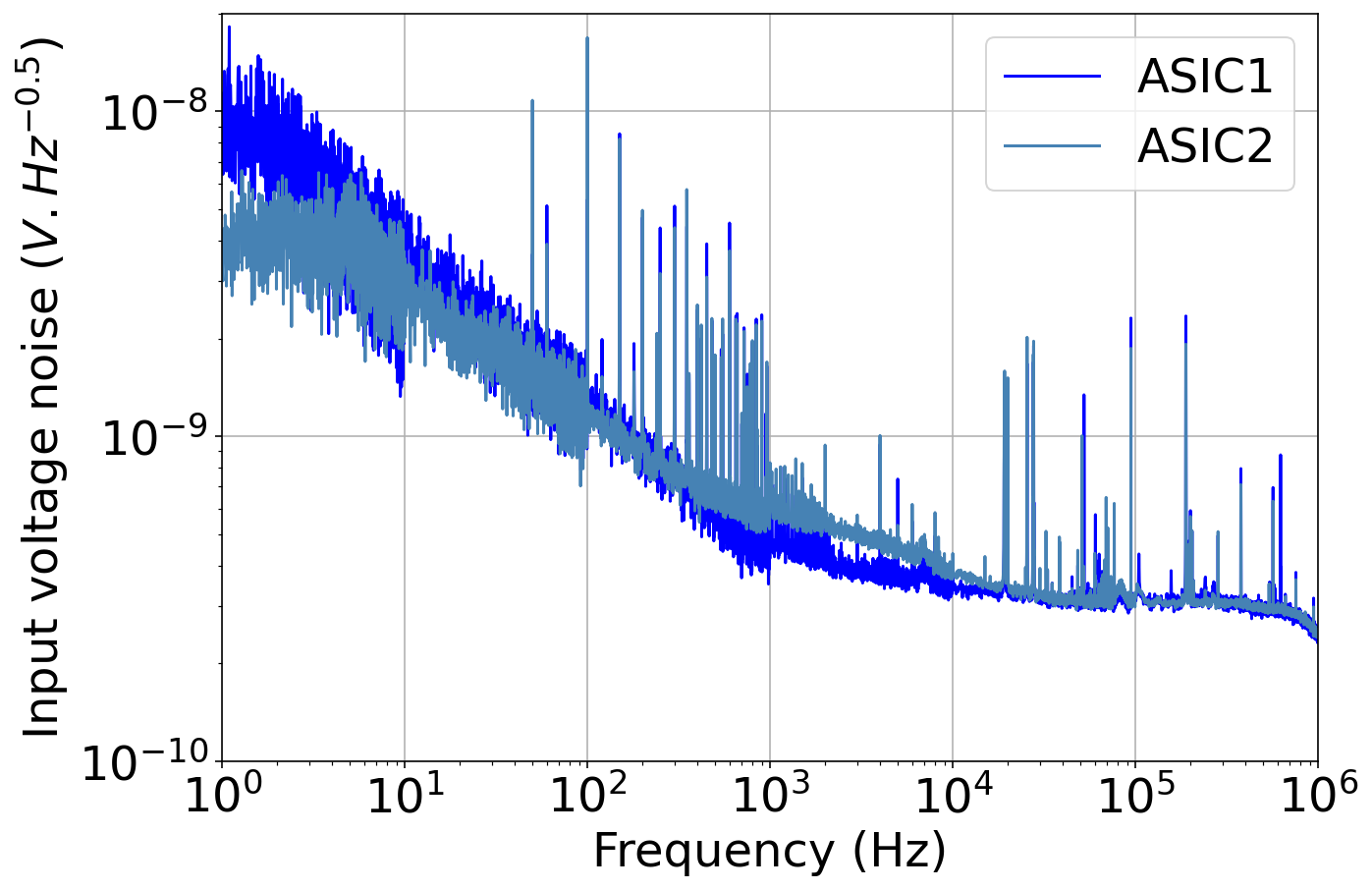}}
\caption{Multiplexed LNA (low noise amplification) equivalent input noise voltage measurement at 300~K and 77~K (top) and at {about 72~K} for the two ASICs in QUBIC (bottom).}
    \label{fig:LNA}
    \end{center}
\end{figure}

From 300~K to 77~K, the voltage gain increases from 20 to 70 as expected and the white noise level decreases from 0.95~nV/$\sqrt{Hz}$ to 0.25~nV/$\sqrt{Hz}$ as shown in Figure \ref{fig:LNA}. The corner frequency also increases from about 100~Hz at room temperature to about 6~kHz at 77~K. The presence of an excess noise below {100~Hz} at low temperature is not understood. The 3~dB bandwidth of the LNA is estimated at about 6~MHz by simulation, not measured precisely because of the limitation from the measurement setup. 

\subsubsection{AC bias current source}
The AC bias current source allows to alternately bias two consecutive SQUIDs of the same row at +$I_{sq}$ and -$I_{sq}$ through addressing capacitors (no power dissipation on the cryogenic stages as compared to the addressing with resistors).
It consists of two differential branches, each of them having an inverter degenerated by current mirrors referenced to the biasing current source described above. These inverters are controlled in phase opposition to the rate imposed by the column changes. Alternately, the outputs of these inverters simultaneously push and pull the same $I_{sq}$ biasing current through each SQUID of the same row through the addressing capacitors. A 1:32 multiplexer located at the output of the inverters of the AC source allows the selection of the row to be biased.
In order to avoid a drift of the operating point at the output of the inverters of the AC biasing circuit and a risk of saturation of the current sources, these  outputs are connected to the voltage reference $V_{init}$ through 2 external resistors of 10~k$\Omega$.

\subsubsection{Multiplexer addressing circuit}
\begin{figure}
 {\includegraphics[width=15cm]{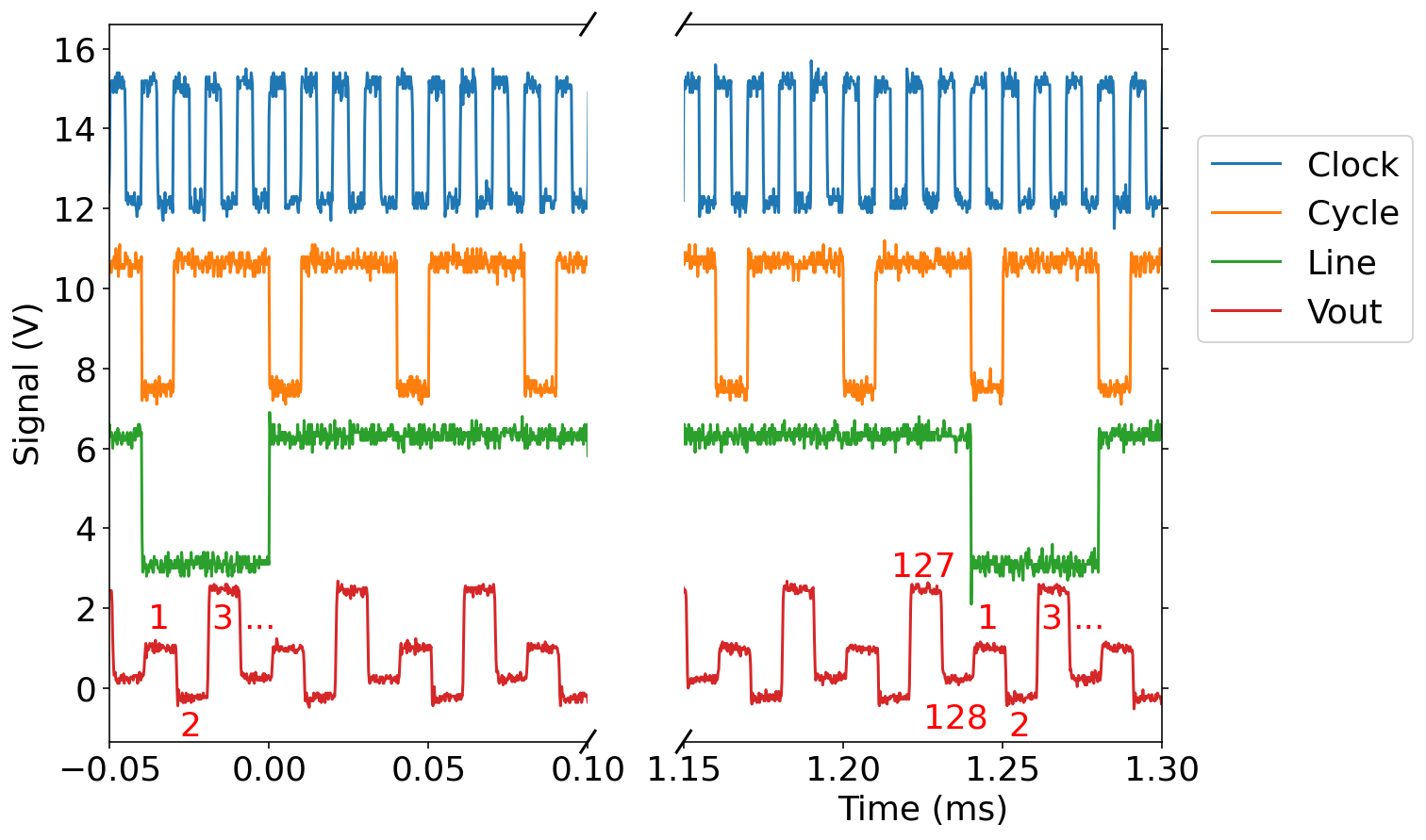}}
\caption{Functional clocking validation at 4.2~K
of the multiplexer: Line: synchronize the SQUID switching current source to the multiplexed LNA; Cycle: 
give the start - pixel 1 - of the full multiplexing cycle; Vout is the multiplexed signal of 128 pixels with the SQUID stage
replaced by 128 resistors biased through capacitors (4 different offsets are noticeable). { The different data have been scaled and shifted for clarity. The numbers in red give the channel ordering.}\label{fig:asicclock}
}\end{figure}

The sequencing of the multiplexing is carried out internally {in} the ASIC by an integrated digital circuit referenced to an external clock signal CK of 100 kHz nominal frequency. In addition to the control of the LNA and the SQUID biasing circuit, it generates two signals active on falling edge, SYNCCb and SYNCLb, that  indicate the end of row and complete cycle respectively. 
%This addressing circuit consists of synchronous counters / down-counters on "CKb" with parallel loading and asynchronous SET / RAZ: it implements a 2-bit counter for the addressing of the columns (multiplexed amplifier inputs selection) followed by a 5-bit counter for the one of the rows (SQUIDs biasing current sources selection):
%- at the end of each 2-bit counter cycle, it generates a "SYNCCb" signal, active on a falling edge, which increments the 5-bit counter (jump to the next row);
%- at the end of each 5-bit counter cycle, it delivers a synchronization signal "SYNCLb" (end of complete cycle) active on falling edge.
%To be consistent with the AC biasing principle of SQUIDs, the addressing mode requires a systematic change of column at the end of each change of row (reading path in "Z" pattern).
%The digital input "COUNTb" is used to define the addressing excursion ("COUNTb" = "0" (default), the array is read from column 1, line 1 to column 4, line 32; "COUNTb" = "1", the array is read from column 4, line 32 to column 1, line 1).
%Addressing is fully reconfigurable and can be adapted to any size included into an array of 4 columns of 32 SQUIDs in series. To do this, each counter has 2 registers: one for the parallel loading of the counting start value (prepositioning), the other for the parallel loading of the counting end value (reset to the value of prepositioning).
The addressing circuit clocked at a multiplexing frequency of 100~kHz was functionally tested down to a temperature of 4.2~K as shown in Figure \ref{fig:asicclock} \cite{2014JLTP..176..433P}.

\subsubsection{Functional tests of the ASIC with SQUIDs}
Functional tests of the ASIC have been performed down to 4.2K in a dedicated cryostat on a small array of 2 columns of 2 SQUIDs in series as shown in Figure \ref{fig:asicclock}. % which consists in 4 “StarCryo” SQUIDs chips bonded on a Printed Circuit Board (PCB) with Surface Mount Device (SMD) addressing capacitors associated to our cryogenic ASIC for the readout and the multiplexing. 
The settle time of the system after switching from one channel to the other is of the order of 5~$\mu$s.
These tests have validated the AC SQUID biasing operation and the overall multiplexing topology (switching AC current sources, multiplexed LNA and digital clocking).

\begin{figure}
    \begin{center}
    {\includegraphics[width=0.9\linewidth]{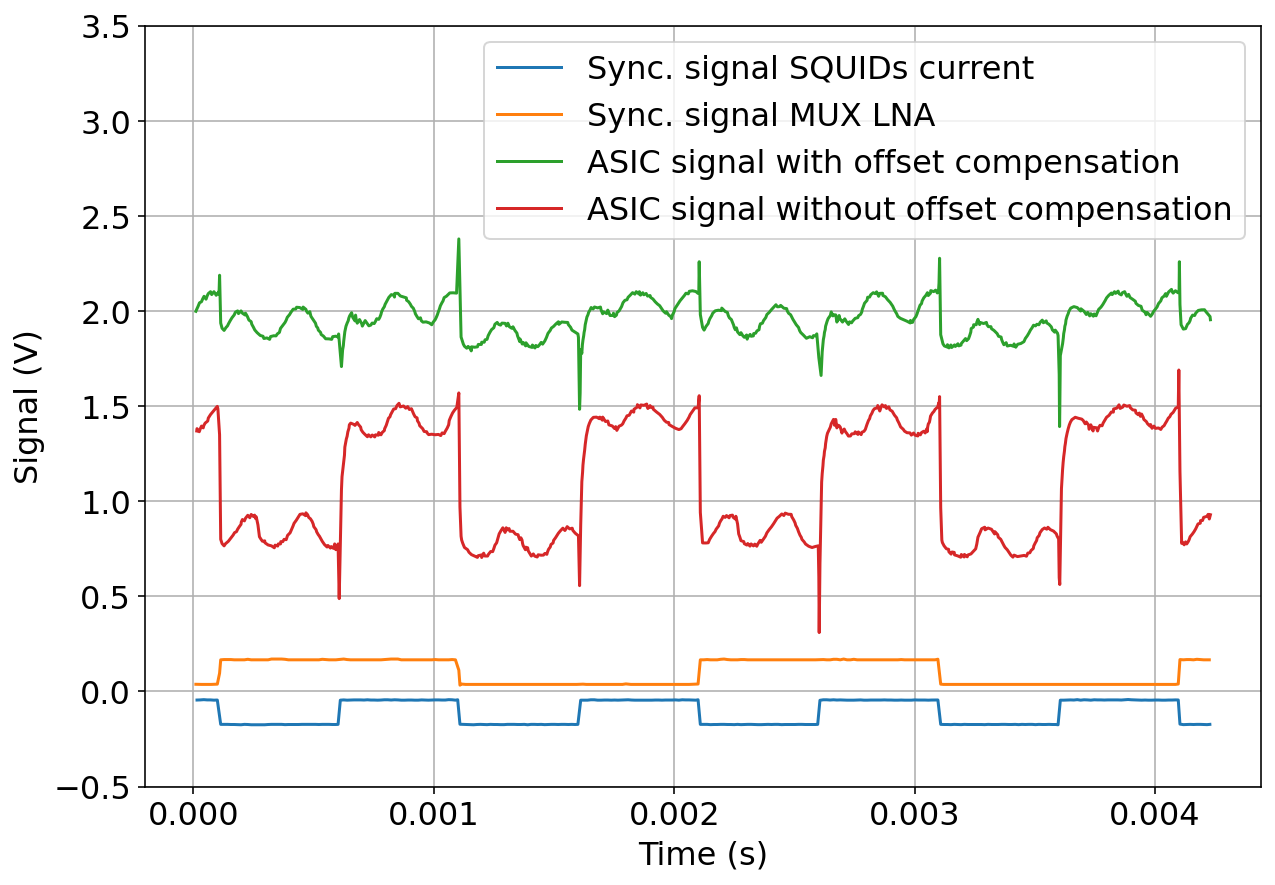}} 
    \caption{Validation at 4.2~K of the ASIC and SQUIDs AC biasing operation through addressing capacitors (100~nF). The tests are performed on an array of 2 columns of 2 SQUIDs in series associated to the cryogenic ASIC. {The measured data have been scaled and shifted for clarity.} Signals in {blue} and {orange} are synchronization signals of the SQUID switching current source and the multiplexed LNA respectively. Signals in {green and red} are the measured multiplexed output signal, with and without offset compensation {respectively}, corresponding to voltage-flux characteristics of each SQUID obtained by applying a large ramp signal into their feedback coil.}
    \label{fig:asicsquidstest}
    \end{center}
\end{figure}

%\subsubsection{Noise level}
%Low noise multiplexed amplifier characterizations have been investigated using a vector analyzer. A white noise level of 0.3 nV/$\sqrt{Hz}$ with a differential voltage gain of 200 and a bandwidth of 6 MHz were measured at 77K, as shown on Figure \ref{fig:LNA}.

%\section{SQUIDs characterization at APC}
%\label{sec:squids}
\subsection{SQUIDs tests and characterizations}
The SQUIDs are first selected with room temperature measurements and furthermore characterized at two temperatures in the QUBIC readout system.

\subsubsection{Selection and sorting of SQUIDs at 300~K}
Before installation in QUBIC, the manufactured SQUIDs undergo a visual inspection in a clean room in order to detect and remove the ones exhibiting evidence of defects during
fabrication or storage. 
We further proceed in the measurement of 4 resistance values at room temperature: 
heater, SQUID washer, feedback inductance and input inductance. 
The distribution of these values is found to be close to a Gaussian with a standard deviation of about 5\% the mean value.
%A histogram of the measured values is plotted in Figure~\ref{fig:SQUIDhistogram}. 
%%The SQUIDS with clearly unacceptable measurements, such as infinite values in one of the 4 resistance values or zero resistance value in leakage test, areimmediately rejected.
%\begin{figure}[ht!]
%    \centering
%    \includegraphics[width=.45\linewidth]{Rfbn}
%    \includegraphics[width=.45\linewidth]{Rhn}\\
%    \includegraphics[width=.45\linewidth]{Rsqn}
%    \includegraphics[width=.45\linewidth]{Rinn}
%    \caption{\label{fig:SQUIDhistogram}Plot of the measured resistance element values of the SQUIDs at room temperature~(Ohms). From left to right : (a)~feedback coil,
%      (b)~heater, (c)~SQUID washer, (d)~input coil}
%\end{figure}
%{As the constructor specifications say, a part of the SQUID is nonfunctional due to the process of fabrication. We also add a more precise range for our SQUIDs to be able to be implemented inside the QUBIC readout system. Some of our tests are just fail or pass process (open circuit or not) others have a range where they are considered as acceptable.\\
%As the distribution of such parameters follows a binomial law, we know that in 3$\,\sigma$ wide we have $98\,\%$ of our detectors that are working. Then we empirically choose to set a border between 2 and 3 $\sigma$ to keep the best SQUIDs to be integrated and in a case where all the best SQUIDs are used or downed, we could accept to work with some SQUIDs that have a lower range of precision (used as spare parts)}\\
SQUIDs with all parameters within 2$\sigma$ of the mean values are selected for installation in
QUBIC.  SQUIDs that are between 2$\sigma$ and 3$\sigma$ for one or more measurements are held aside as a possible option in case there are not enough SQUIDs passing the
first criteria.  All SQUIDs with any parameter larger than 3$\sigma$ from the mean are rejected. 
While these room temperature measurements do not guarantee that a SQUID is functional, the 2 and 3~$\sigma$ selection process has been chosen as a trade-off between the number of available chips and the expected homogeneity in the SQUID behaviour.
%The SQUIDs are furthermore subjected to washer leakage tests to measure the resistance to the input and to the feedback inductance. 
A further selection process is performed based on the leakage resistance between SQUID washer and the input inductance.  
Leakage measured at cryogenic temperature is typically a few M$\Omega$ between a full
stack of 32~SQUIDs and the 32~input inductances.  This level of leakage
does not significantly degrade the operation of the SQUIDs.  The pass/fail
level for leakage to the input inductance was therefore set at 2~M$\Omega$, with the majority
of leakage values measured closer to 20~M$\Omega$.  SQUIDs with
leakage to the input inductance less than 2~M$\Omega$ were rejected in order to avoid any risk of electrostatic discharge damage. 
For the same reason, the leakage between the SQUID washer and the feedback must be that of an open circuit
($\mathrm{resistance}>40~\mathrm{M}\Omega$), otherwise the SQUID is rejected.
%The measurements were done at the highest resistance range of the multimeter in order to limit the current going through the SQUID.
% All tests were carried out with an Amprobe multimeter, and the reference values defined by it.
We typically obtained a yield of about 82\% for tested SQUIDs.% package of $16\times16$.

\subsubsection{Tests at Cryogenic Temperature}
The characterization of the SQUIDs is performed at the begining of the calibration phase, with the focal plane temperature kept just above the TES critical temperature in order to be in normal state and to reduce the detector current noise contribution. The main goal is to define the optimal SQUID bias current to be used during observations. 
%Measurements were performed with the QUBIC Technical Demonstrator.
%TES detector array P87 was mounted in the focal plane.  The instrument
%had the window blocked outside the cryostate by a metal sheet, at
%approximately 300~K.  The TES were at a temperature of 378~mK, while
%the SQUIDs were at a temperature of 4.146~K, and the pressure in the
%cryostat was~$4\times10^{-6}$~mbar.

\begin{figure}[ht!]
    \centering
    {\includegraphics[width=0.8\linewidth]{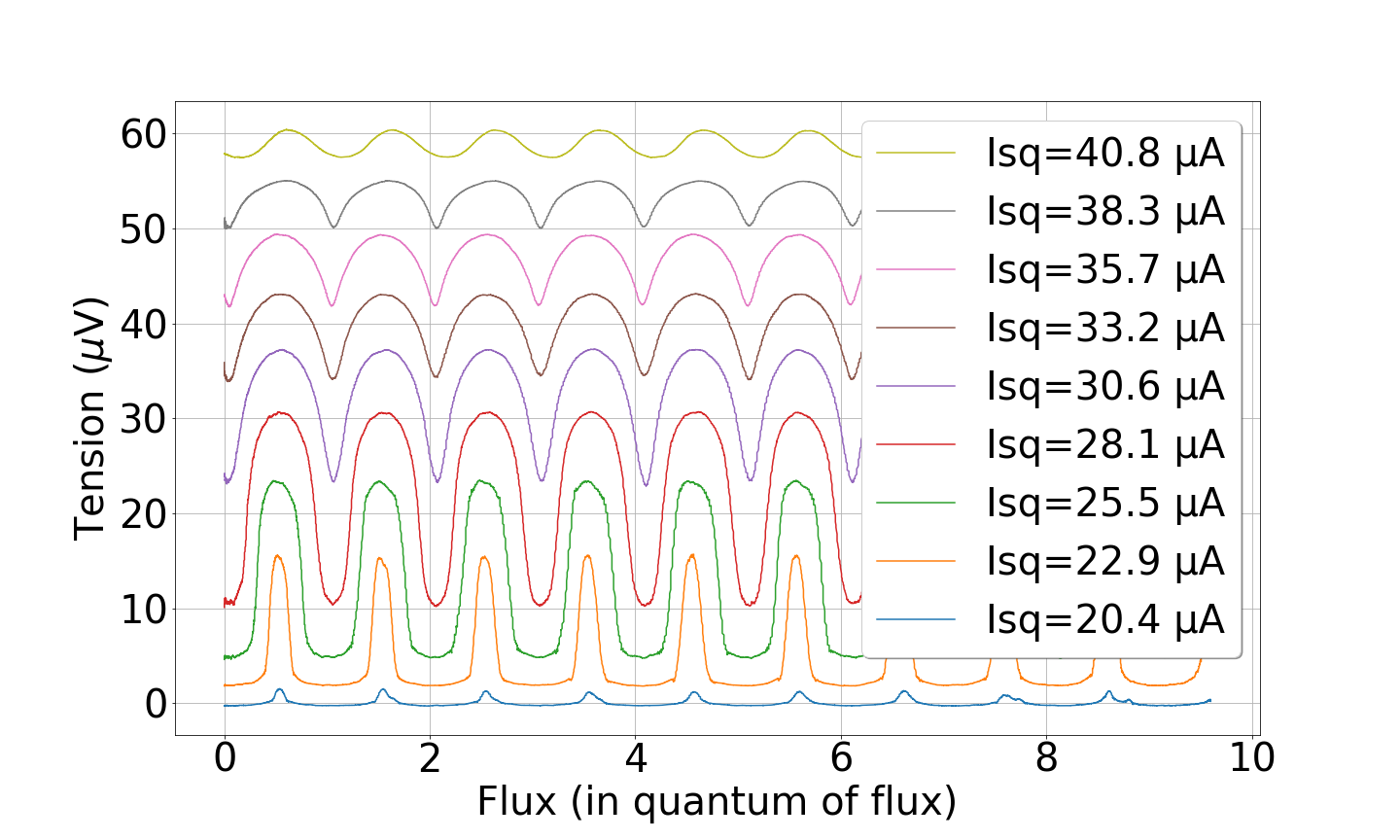}}
    \caption{\label{fig:fluxSQUID1}Flux-to-voltage SQUID transfer
      function for current biasing on ASIC~1. The plots show the response signal ($V_\mathrm{sq}$) as a function of the quantum flux going through the SQUID.  There are 9~curves corresponding to increasing bias current ($I_\mathrm{sq}$) from bottom to top.}
\end{figure}

The principle of the procedure is the following: a slow sinusoidal signal of 12~seconds period and 1~V peak-to-peak
amplitude is injected into the feedback inductance through the feedback resistor $R_\mathrm{fb}$=10~k$\Omega$ and the bias current of the SQUIDs is increased step by step.
For each value of the input
current $I_\mathrm{sq}$, the response of the SQUID is therefore measured as shown in Figure~\ref{fig:fluxSQUID1}.
%Simultaneously, it is determined the current which results in the optimal response bias. 
%Figure~\ref{fig:fluxSQUID1} show examples of
%the response of the SQUID to 9~settings of the $I_\mathrm{sq}$ from ASIC~1 and ASIC~2.

\begin{figure}[ht!]
    \centering
    \includegraphics[width=.49\linewidth]{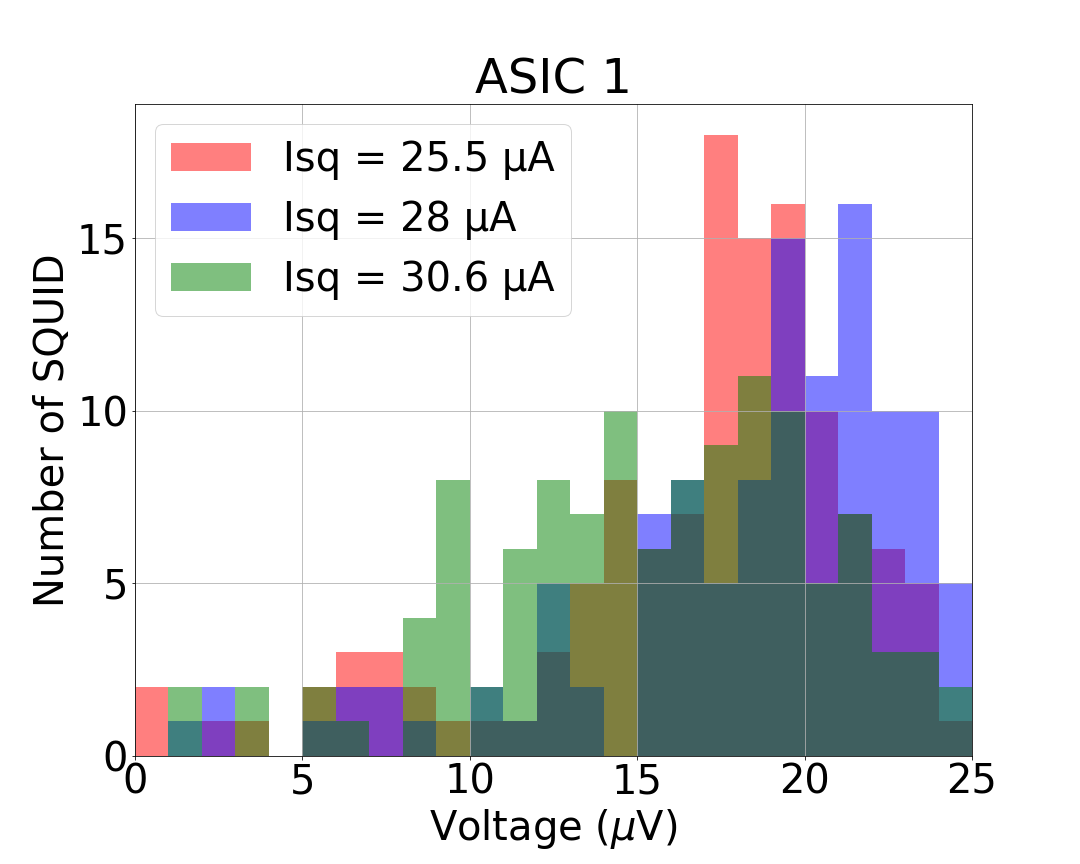}
    \includegraphics[width=.49\linewidth]{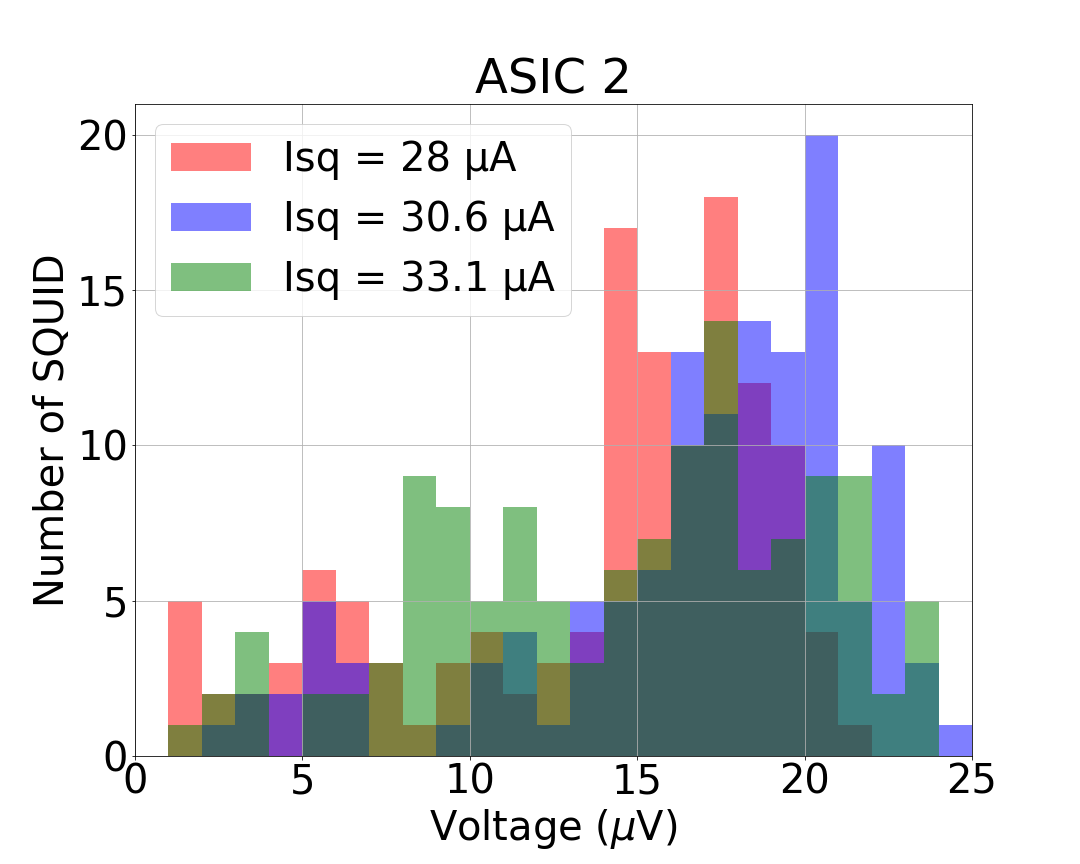}
    \includegraphics[width=.49\linewidth]{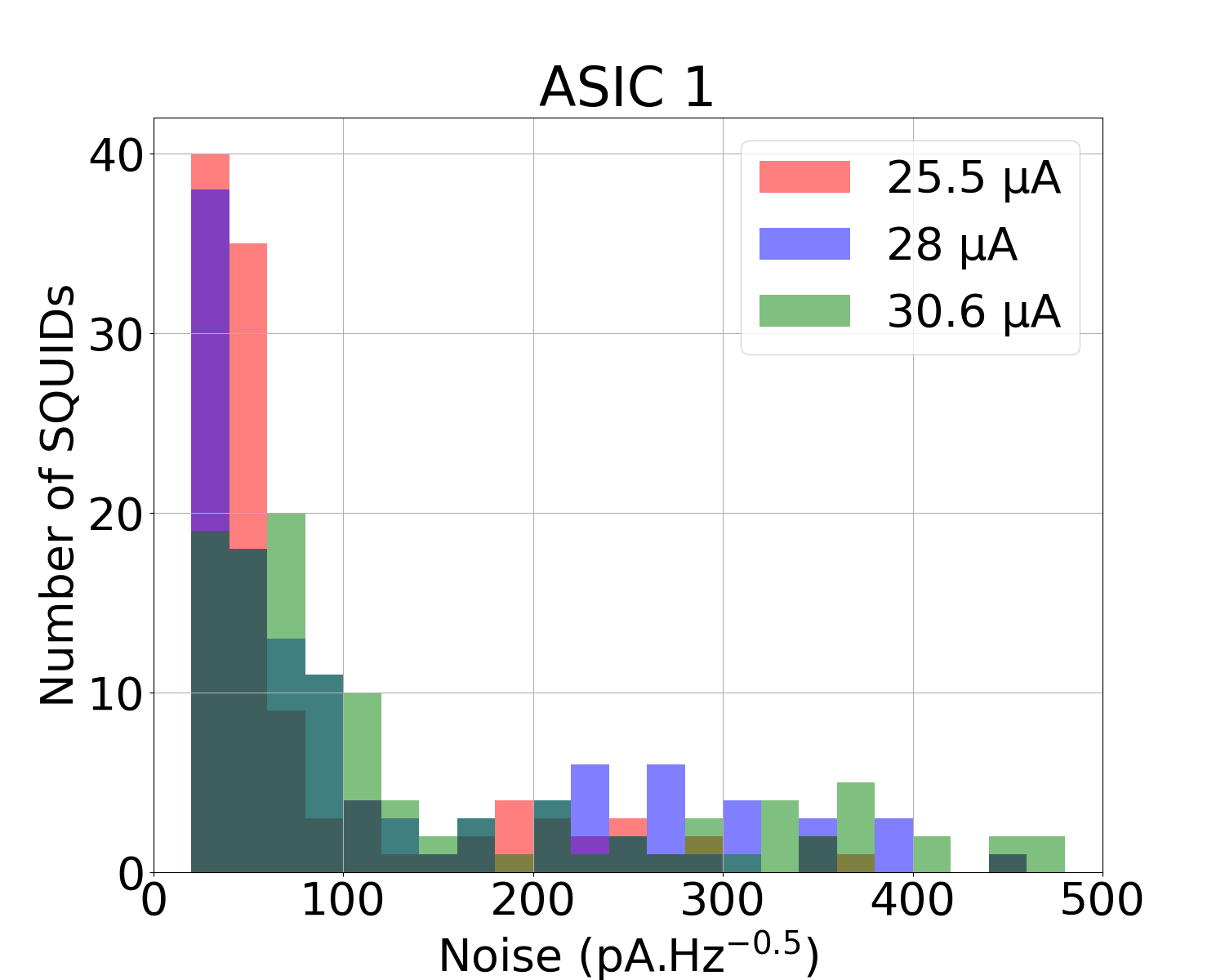}
    \includegraphics[width=.49\linewidth]{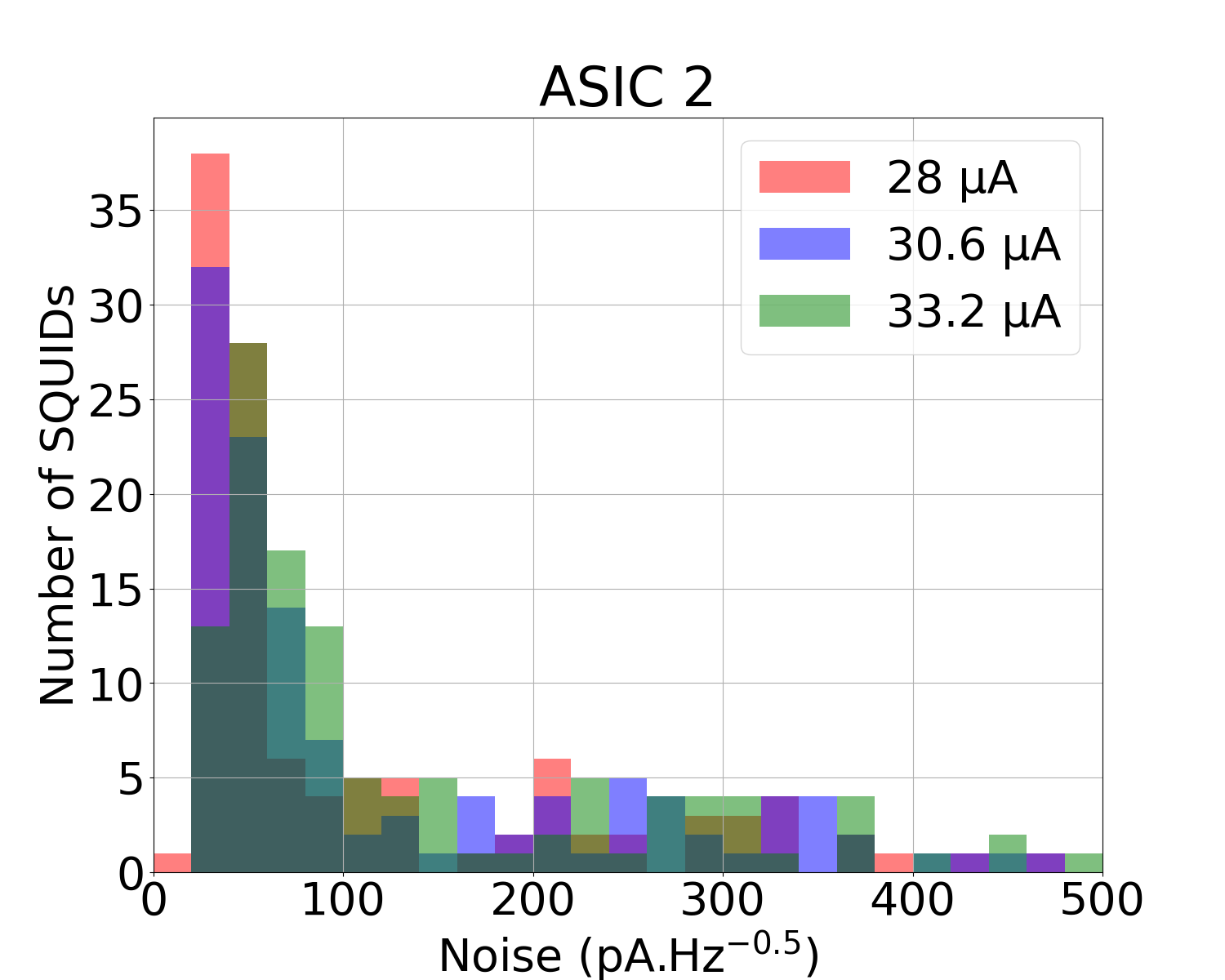}
    \caption{\label{fig:SQUIDcurrenthisto1}Histograms of SQUID voltage swing (\textit{top}) and current noise (\textit{bottom}) for SQUIDs connected to ASIC 1 (\textit{left}) and 2 (\textit{right}), for three bias currents around the optimal one.}
\end{figure}

%The bias current steps through 16~indexed values from 0~to~15, but the
%lower 7~values produce undetectable response from the SQUID.  Only the
%final 9~values are plotted.
As the SQUID current $I_\mathrm{sq}$ increases, the amplitude of the response of
the SQUID also increases until it reaches a maximum and then it decreases.  The
optimum $I_\mathrm{sq}$ corresponds to the maximum amplitude of the
SQUID response.
Since the same $I_\mathrm{sq}$ must be supplied to all the SQUIDs per
ASIC, it is necessary to select a single bias index for all the SQUIDs
for each ASIC.  
%Figure~\ref{fig:SQUIDbiashisto1} shows the histogram of optimal bias for ASIC~1 and  ASIC~2.
%\begin{figure}[ht!]
%    \centering
%    \includegraphics[width=0.45\linewidth]{ASIC1_Histogram}
%    \includegraphics[width=0.45\linewidth]{ASIC2_Histogram}
%    \caption{\label{fig:SQUIDbiashisto1}Histogram of the optimum biasing current for each SQUID in ASIC~1 (\textit{Left}) and in ASIC~2 (\textit{Right}).}
%\end{figure}
%The best bias index is one of the three near the peak of the histogram.  
While it seems natural to choose the SQUID current bias corresponding to the majority of the SQUIDs, % of the the peak of the histogram,
in reality it does not maximize the number of operational SQUIDs. 
%not all the SQUIDs are considered operational at that current bias.
A SQUID is considered operational if its response is
greater than 10~$\mu$V.  The SQUID current is  therefore chosen to maximize the number of
operational SQUIDs. We also deduced from these data the current noise by dividing the voltage noise (averaged between 40~Hz and 50~Hz) by the slope in the middle of the flux-to-voltage transfer function and  by the input coil mutual inductance.
Figure~\ref{fig:SQUIDcurrenthisto1} shows the histograms of the SQUID response and the deduced current noise for three
$I_\mathrm{sq}$ bias current for the two ASICs. 

The SQUID voltage swing histograms of Figure~\ref{fig:SQUIDcurrenthisto1} show that 28~$\mu$A is the optimal bias current for ASIC~1 for which 93\%
of the SQUIDs are operational.  For ASIC~2, the histograms give 30.6~$\mu$A as the best bias
current with 91.1\% operational SQUIDs. In terms of current noise, the distribution is slightly more peaked for these optimal bias current as shown in the bottom histograms of Figure~\ref{fig:SQUIDcurrenthisto1}, with a median value around 70~$pA/\sqrt{Hz}$ dominated by the TES aliased current noise (see section \ref{subsect:noise}).

%Figure~\ref{fig:SQUIDfocalplane} shows the distribution of the SQUID
%bias index on the focal plane of the QUBIC Technical Demonstrator.
%\begin{figure}[ht!]
%   \centering
%    \includegraphics[width=0.9\linewidth]{carto_P87}
%    \caption{\label{fig:SQUIDfocalplane}The QUBIC Technical
%      Demonstrator quarter focal plane showing the optimal
%      $I_\mathrm{sq}$ index as a colour code.}
%\end{figure}

The yield of SQUIDs for the QUBIC TD is
93\% for the 128~SQUIDs connected to ASIC~1, and 89\% for the
128~SQUIDs connected to ASIC~2.  This corresponds to 119~operational
SQUIDs for ASIC~1 and 114~operational SQUIDs for ASIC~2.  The total
yield is therefore 91\%.  The optimum bias current is 28.1~$\mu$A for
ASIC~1 and 30.6~$\mu$A for ASIC~2.

%\clearpage
\section{TES characterization}
%\section{TES characterization in the APC Dilution Cryostat}
\label{sec:TEScharacterization_functionality}
The TES array currently used in QUBIC has the reference P87 of the production series. It has been selected after characterizations both at room and cryogenic temperatures. This array has been also fully characterized during the QUBIC calibration phase. 
%and all this process will be illustrated with this array.

\subsection{Selection process and integration}
The TES arrays that have been successfully produced undergo visual inspections and resistance measurements at room temperature on a probe station. These measurements are done before integration and wire bonding to detect possible defects or issues with the routing. If successful, the array is integrated in the QUBIC holder and Al wire bonded  (Figure~\ref{testransi}).

The next steps consist in characterizations at cryogenic temperature. They are done in an Oxford Instrument dilution fridge before integration in the QUBIC cryostat. 

%Characterization of the TES array ``P87'' was first done using an Oxford dilution fridge in the APC Millimetre Wave Lab.
% shown in Figure~\ref{fig:dilu}.
%A specific thermal architecture has been done in order to reproduce the QUBIC This system is capable of cooling the test device down to temperature
%of 20~mK.  Using a feedback system of resistance heaters and
%temperature sensors, the Oxford Instruments cryogenic control system
%permits temperature stability at a user defined temperature within the
%range of 20~mK and 500~mK or more.

%\begin{figure}[ht]
% \centering
% \includegraphics[width=0.9\linewidth]{TESarray_testcryo2}
% \caption{Photograph of the dilution refrigerator system in the APC
%   Millimetre Wave Lab.\label{fig:dilu}}
%\end{figure}

%Figure~\ref{fig:TES_FP_dilution} shows the I-V curves for 244 TES
%bolometers which lie in the focal plane of the QUBIC Technical
%Demonstrator.  Four TES are outside the focal plane (not shown) and
%are used as dark detectors for comparison.

\subsection{Critical Temperature}
\begin{figure}[ht]
\centering
{\includegraphics[width=0.9\linewidth]{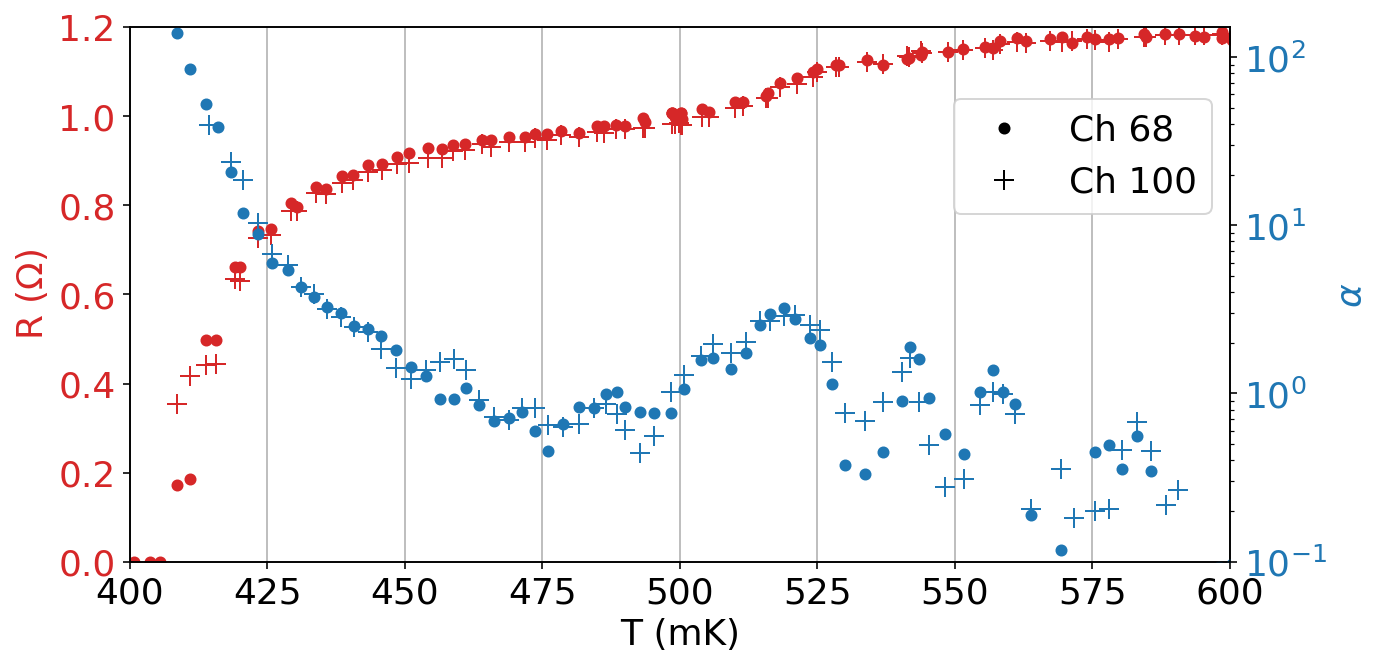}}
\caption{Resistance as a function of temperature for array P87 dark pixels on channels 68 and 100 (red) and derived $\alpha = \frac{T}{R}\frac{\partial{R}}{\partial{T}}$ parameter  (blue).\label{fig:Tcritical0}}
\end{figure}

The QUBIC detector wafer includes 8~dark pixels, 4 for each ASIC besides the 124 active ones. These channels consist in NbSi thermal sensors of the same geometry as the ones used on TESs, without thermal decoupling from the silicon wafer. They are located outside the focal plane and are therefore shielded from radiation. 
Figure~\ref{fig:Tcritical0} shows measurements
of the transition from normal to superconducting state for 2 of these dark pixels, measured in a dilution fridge cryostat (which is a dedicated test bed for selection of detectors) by increasing slowly the temperature and with a QUBIC readout chain. The critical temperature is about 412~mK and some temperature dependence is still present above the transition, which allows to still have some sensitivity in case of saturation. The unit-less parameter $\alpha=\frac{T}{R}\times \frac{\partial{R}}{\partial{T}}$ has been evaluated from these transition curves and is higher than 100 for the lowest part in the transition. A small transition is visible at about 520~mK in this P87 array which is not understood.

\subsection{TES normal and parasitic resistances}
\begin{figure}[ht]
 \centering
 {\includegraphics[width=0.47\linewidth]{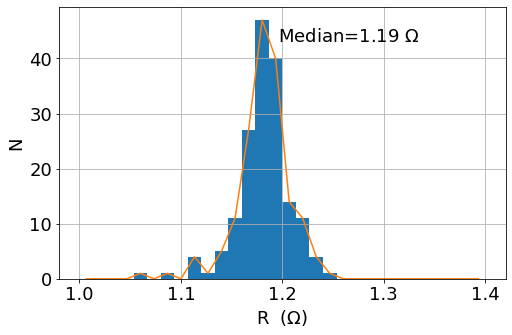}
 \includegraphics[width=0.47\linewidth]{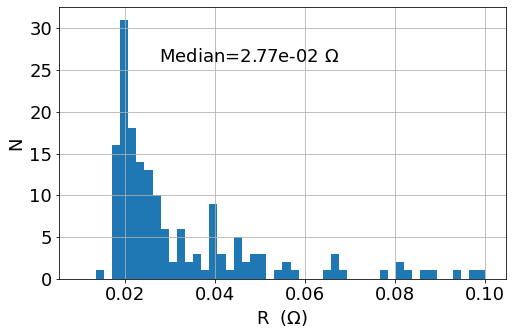}}
\caption{Histogram of normal resistance ({left}) and of residual resistance in the superconducting state ({right},
  {167 total number of TES for both graphs}). This residual resistance is the sum of a
  parasitic resistance and the bias resistor of 10~m$\Omega$.
   \label{HistoRnormalsupra}}
\end{figure}

With the bath temperature below the TES critical one, the detectors need to be over-biased (above about 7~$\mu$V) in order to be in the normal state.
A slow and small sine wave voltage oscillation was added in order to
deduce the resistance value. 
Figure~\ref{HistoRnormalsupra} top shows the distribution of the normal resistance values for the array P87. It is highly peaked around 1.2~$\Omega$ as expected from the transition measurement.

The same procedure is used to determine the resistance in superconducting state, but without any DC bias on the detectors. The residual resistance obtained from these measurements, assuming the TES
resistance is 0~$\Omega$, is given by the sum of the shunt resistance
(10~m$\Omega$) and the parasitic resistance which is in series with
the TES. The parasitic resistance is assumed to come from the connectors used. Figure~\ref{HistoRnormalsupra} bottom shows the distribution of these residual resistance values for the array P87.  The median is {28~m$\Omega$} which leads to a parasitic resistance of about {18~$m\Omega$} compatible with previous measurements on a dedicated test bench.

\subsection{TES parameters}
\begin{figure}[ht]
 \centering
 {\includegraphics[width=0.9\linewidth,trim={4cm 4cm 4cm 4cm}, clip]{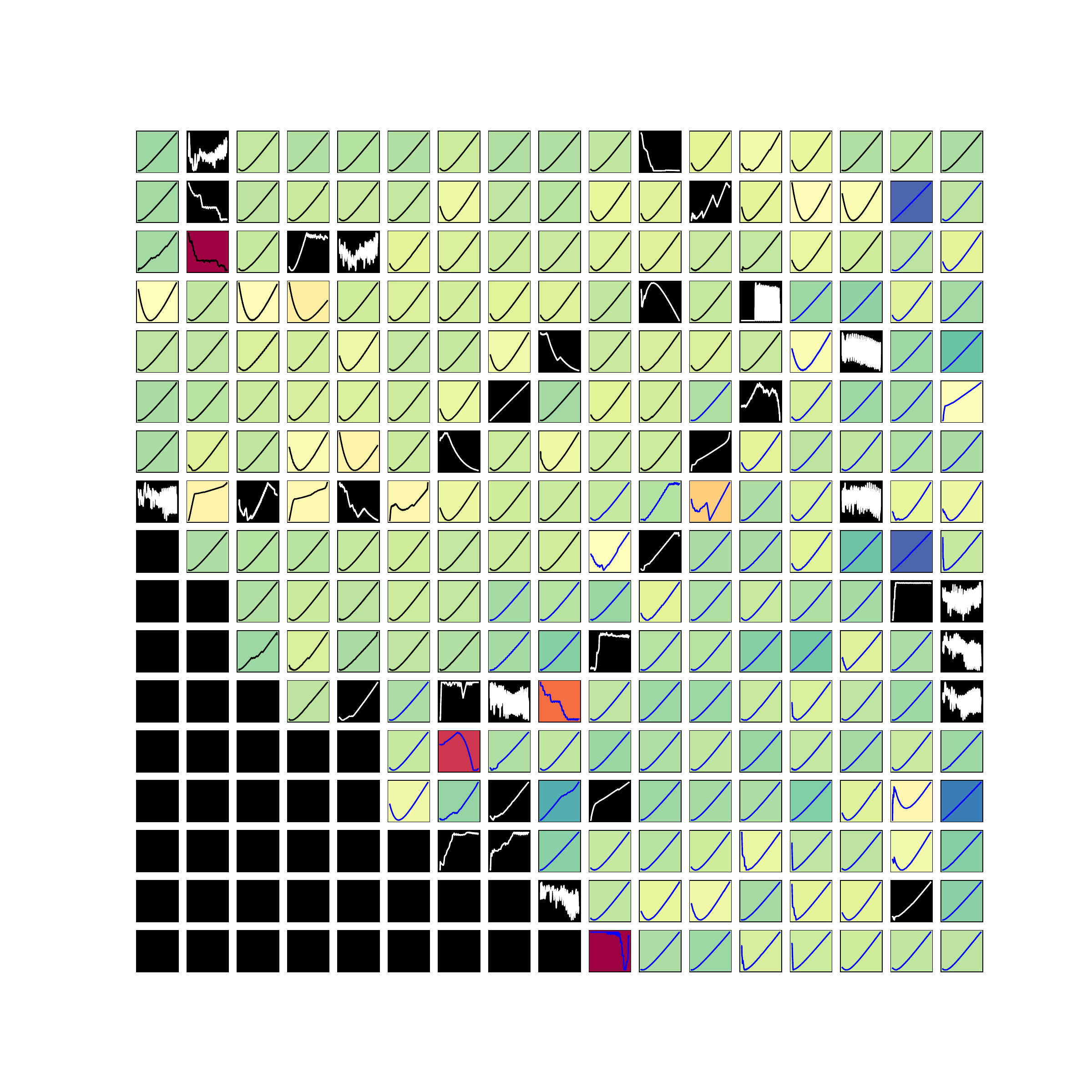}}
 %{QUBIC_focal_plane_20180227T213235}
 \caption{I-V curves for the array of detectors at 360~mK. Each box in the plot shows the measured I-V curve for the detector in that position in the focal plane. {Detailed I-V curves are shown in figure~\ref{fig:TES063-IV-ASIC2}.} This is array P87 measured in the APC dilution cryostat. The vertical axis for each plot is in arbitrary current units, scaled for the minimum and maximum of each plot. There are 244 TES bolometers in the focal plane of
   the QUBIC Technical Demonstrator.  Eight TES are outside the focal plane (not shown) and are used as dark detectors for
   comparison. The background colour indicates the bias
   voltage turnover point.% which is the optimal setting.  
   We see homogeneous characteristics of the TES array and a yield of
   84\% (proportion of TES showing an Electro-Thermal Feedback effect in the I-V curve). The black, filled-in “pixels” in the bottom-left are empty positions.  The QUBIC-FI will have four arrays equivalent to this one in order to make a roughly circular focal plane for each frequency channel.}\label{fig:TES_FP_dilution}
 \end{figure}

\begin{figure}[ht]
  \centering
 {\includegraphics[width=0.9\linewidth]{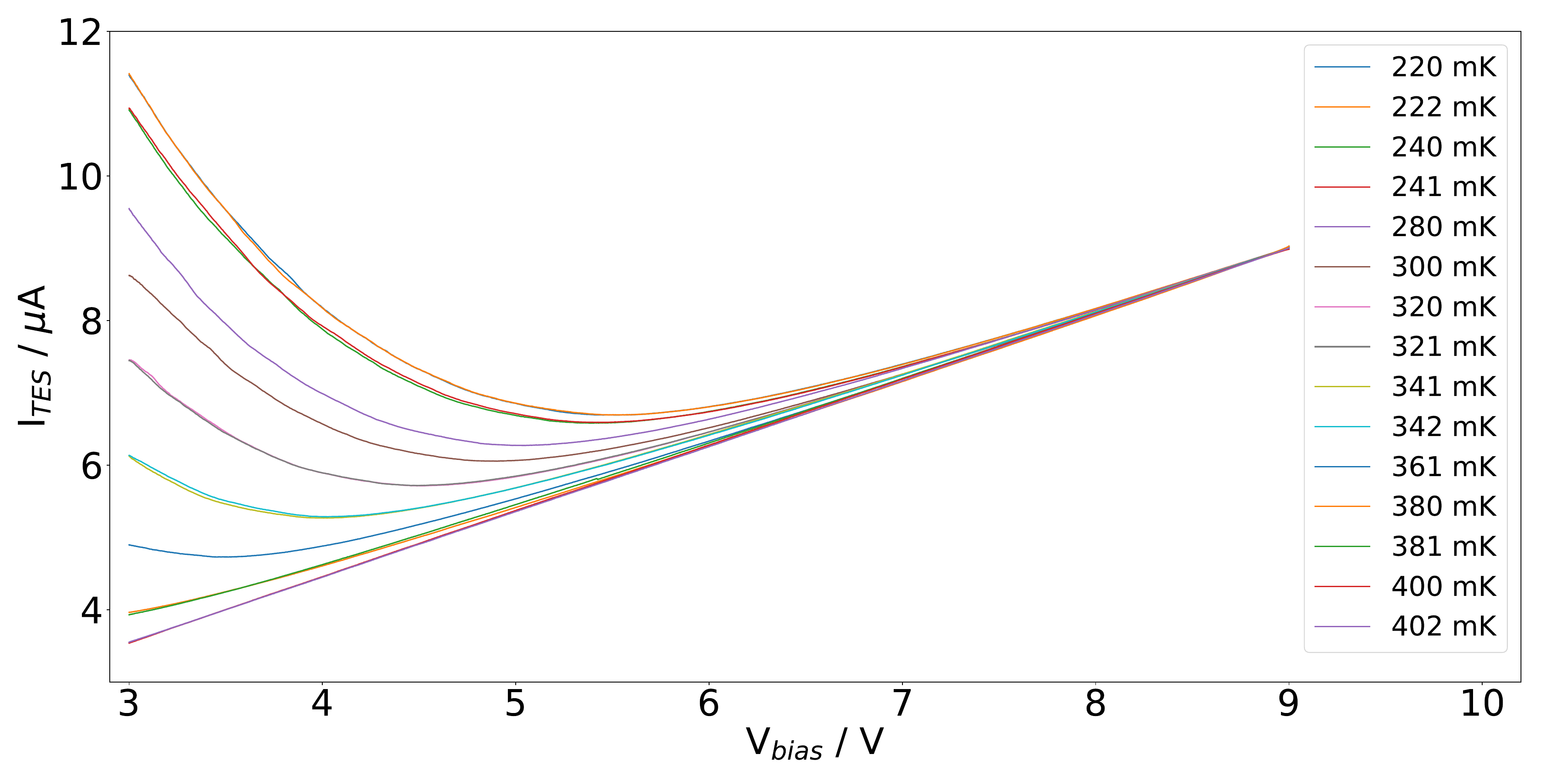}}
 \caption{I-V curves of TES\#63 on ASIC2 at different temperatures. {The TES voltage is obtained from the bias voltage with a $10^{-6}$ divider bridge.}
   \label{fig:TES063-IV-ASIC2}}
 \end{figure}

The I-V characteristics at different
temperatures have been acquired both in a dilution fridge cryostat
%(which is a dedicated test bed for selection of detectors)} 
and in the QUBIC cryostat with optical window open and closed. The measurement follows the procedure outlined in \cite{PerbostPhD}. Figure~\ref{fig:TES_FP_dilution} shows the I-V curves for the P87 array measured at 360~mK in blind configuration and Figure~\ref{fig:TES063-IV-ASIC2} is an example of the I-V curves for one TES on ASIC2 at different temperatures. The strong Electro-Thermal Feedback (ETF) regime is clearly seen with the increase of the TES current at low bias voltages. An overall yield of about 84\% is furthermore obtained in this I-V characterization.

The physical parameters of each TES can be deduced from these measurements. Assuming the TES is in the strong ETF regime and that it is blind, the Power-Temperature relation is classically given by:
\begin{equation}
  P_\mathrm{bias} = \kappa(T_\mathrm{bath}^n - T_c^n)
  \label{eq:PowerTemperature}
\end{equation}
where $P_\mathrm{bias}$ is the bias power dissipated in the TES, $T_\mathrm{bath}$ is the bath temperature, $T_c$ the TES critical temperature, $\kappa$ and $n$ are constants that depend on the thermal link between the absorber and the  bath. In the ETF regime, the bias power is therefore constant for a given bath temperature. The dynamic thermal conductance {G} is further given by the following equation:
\begin{equation}
  G = n\kappa T_c^{n-1}
    \label{eq:thermal conductance}
\end{equation}

\begin{figure}[ht]
  \centering
 {\includegraphics[width=0.9\linewidth]{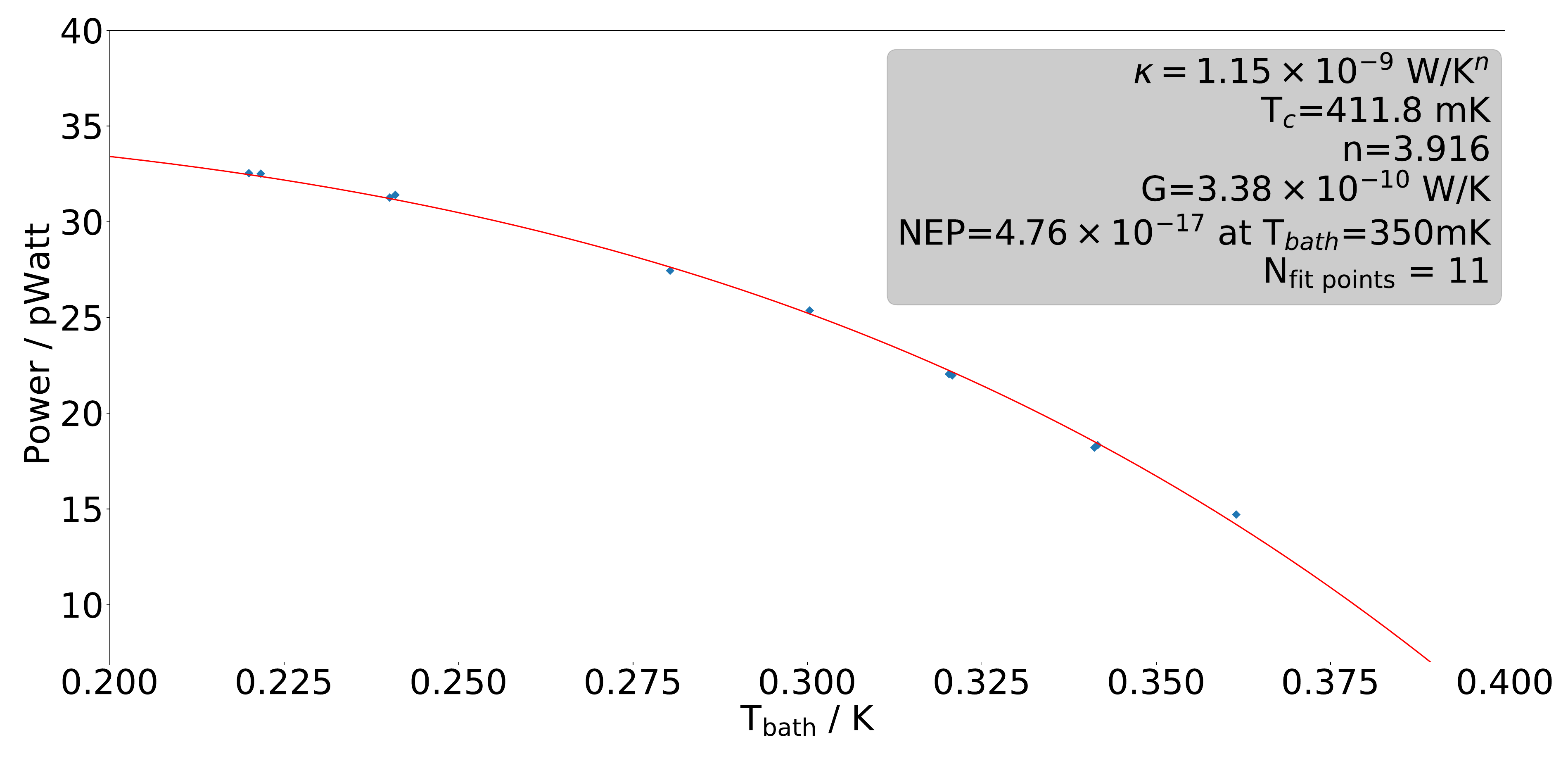}}
 \caption{An example of the fit to the {power-temperature} measurements. This is for TES\#63 on ASIC2. {The best fit parameters from equation \ref{eq:PowerTemperature} are also given, with deduced dynamic thermal conductance and phonon NEP.}
   %The NEP is $4.51\times10^{-17}\mathrm{W}/\sqrt{\mathrm{Hz}}$.
   \label{fig:TES063-NEP-ASIC2}}
 \end{figure}

Figure~\ref{fig:TES063-NEP-ASIC2} shows an example of the
Power-Temperature relation for the same TES as in Figure~\ref{fig:TES063-IV-ASIC2}. A curve fitting algorithm based on Eq.~\ref{eq:PowerTemperature} is used to derive the values of $\kappa$, $n$ and $T_c$ from measured temperatures and powers. 

\begin{figure}[ht!]
    \centering
    {\includegraphics[width=.7\linewidth]{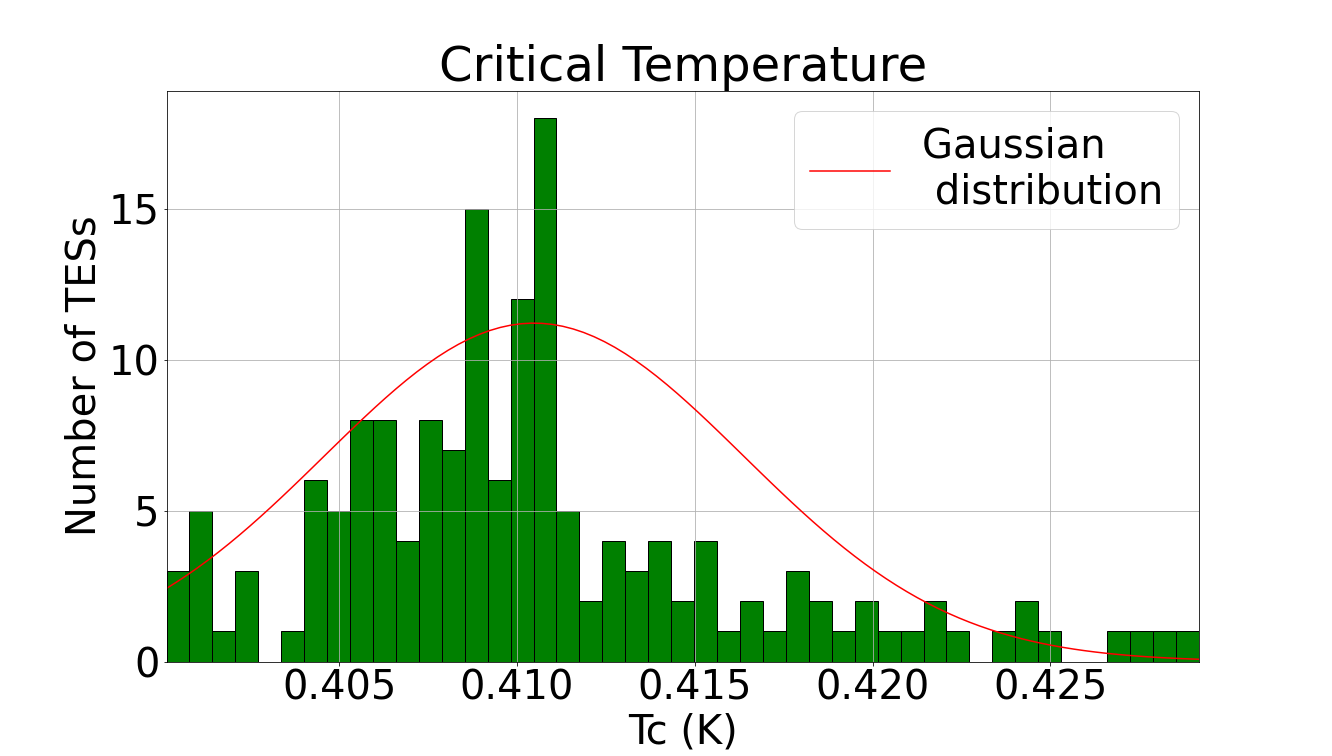}}
    {\includegraphics[width=.7\linewidth]{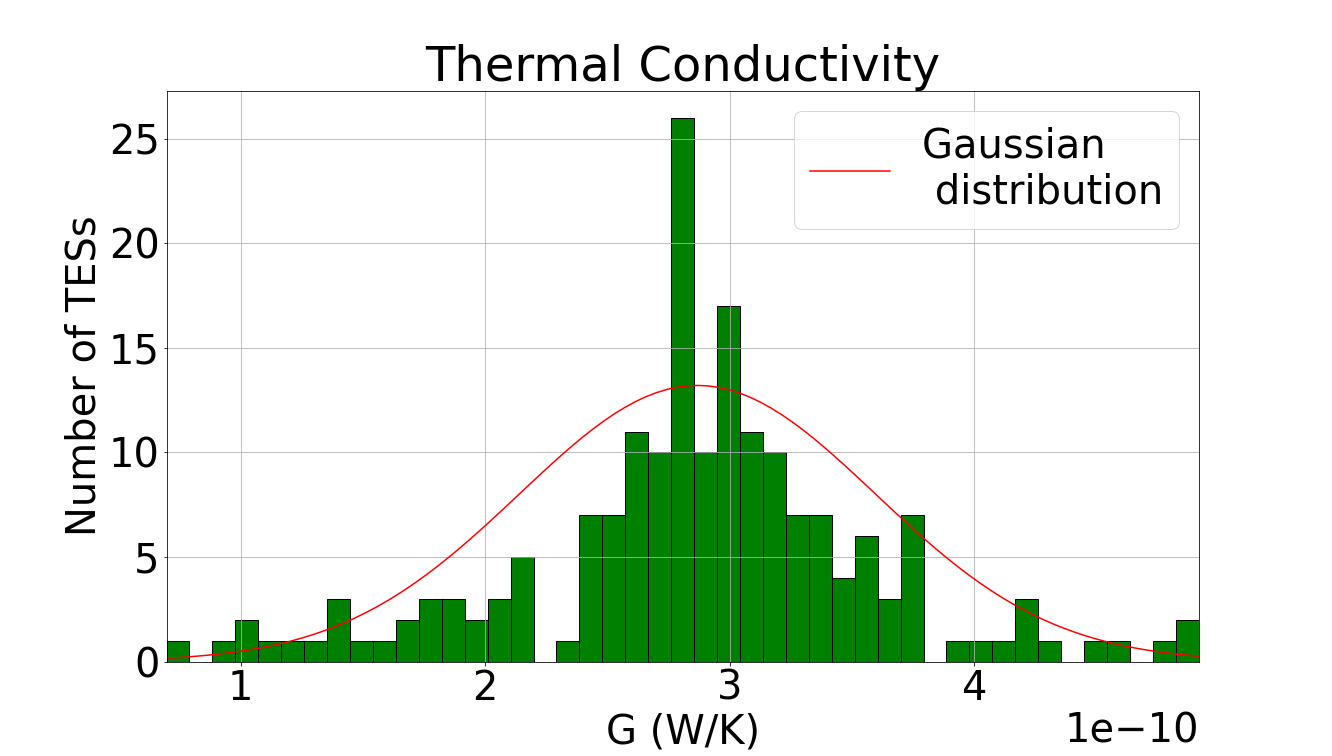}}
    \caption{\label{fig:TESparams} Distribution of critical temperature (top) and dynamic thermal conductance (bottom) of P87 TES array.}
\end{figure}

While degenerate with $\kappa$ as shown in \cite{2018SPIE10708E..45S}, the index $n$ of the power law is around 4 as expected for 500~nm thickness Si$_3$N$_4$ legs.   Figure~\ref{fig:TESparams} shows the distribution of the critical temperature and dynamic thermal conductance obtained with the fit. The critical temperature is around 410~mK as measured on the dark pixels and the median dynamic thermal conductance is about 300~pW/K. The spread in these parameters is probably inherent to the previously quoted degeneracy between parameters in the fit.

\subsection{Detector biasing}
A common bias voltage is used for all 128 TESs readout by one ASIC. As seen in Figure \ref{fig:TES_FP_dilution}, there are some inhomogeneity in the pixel behavior, especially below the turnover, which could leads to over or under biasing some pixels. Going deeper in the transition should wipe out this effect since the responsivity depends only on the bias voltage in strong ETF. We nevertheless experienced some instability at low bias due to the fact that the FLL is no more fast enough with respect to TES time constant. The yield therefore decreases when going deeper in the transition. As a consequence, an optimum has to be found between stability and responsivity, which is usually between 2 and 3~$\mu$V.

\subsection{Power background}
\begin{figure}[ht]
 \centering {\includegraphics[width=0.49\linewidth]{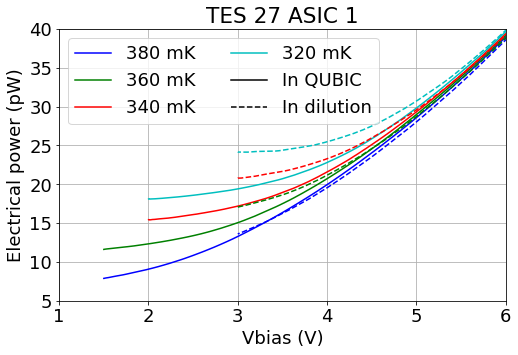}
 \includegraphics[width=0.49\linewidth]{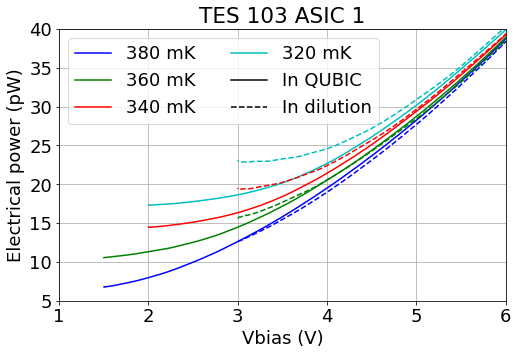}}
 \caption{Examples of electrical power versus bias voltage
   measured in the dilution and in QUBC for two detectors. Comparing the
   electrical power at the same bath temperature in the
   Electro-Thermal Feedback mode (at low bias voltage) gives
   an estimation of the background power.
   \label{PVcomp}}
\end{figure}

The P-V curves measured during blind characterizations and with the QUBIC optical window open are compared in Figure~\ref{PVcomp}. The comparison leads to an estimate
of the power background of the order of a 5~pW which is higher than the expected 1-2~pW from the photometric model of the instrument.
%since there is a Neutral Density Filter with about 9\% transmission at the entrance of the 1~K box. 
This could be due to a difference in temperature sensor calibration between the cryostat used for blind characterizations and QUBIC.

\subsection{Phonon Noise Equivalent Power}
The expected Phonon Noise Equivalent Power
(NEP$_\mathrm{phonon}$) was derived from the fitted parameters with the relation \cite{1982ApOpt..21.1125M}:
\begin{equation}
  NEP_\mathrm{phonon} = \sqrt{\gamma 4k_BT^2G}
  \label{eq:NEP}
\end{equation}
where $k_B$ is the Boltzmann constant, {$T$ is the bolometer temperature} and $\gamma$ is a correction coefficient given by:
\begin{equation}
  \gamma = \frac{n}{2n+1}  \frac{1-(\frac{T_\mathrm{bath}}{T})^{2n+1}}{1-(\frac{T_\mathrm{bath}}{T})^n}.
  \label{eq:gamma}
\end{equation}

\begin{figure}[ht]
 \centering
% \noindent\includegraphics[width=0.45\linewidth,clip]{QUBIC_TES_ASIC1_NEP_histogram}
% \includegraphics[width=0.45\linewidth,clip]{QUBIC_TES_ASIC2_NEP_histogram}\\
 %\includegraphics[trim=0cm 0cm 0cm 2cm, clip=true, width=0.9\linewidth,clip]{QUBIC_TES_NEP_histogram}
 {\includegraphics[width=0.9\linewidth]{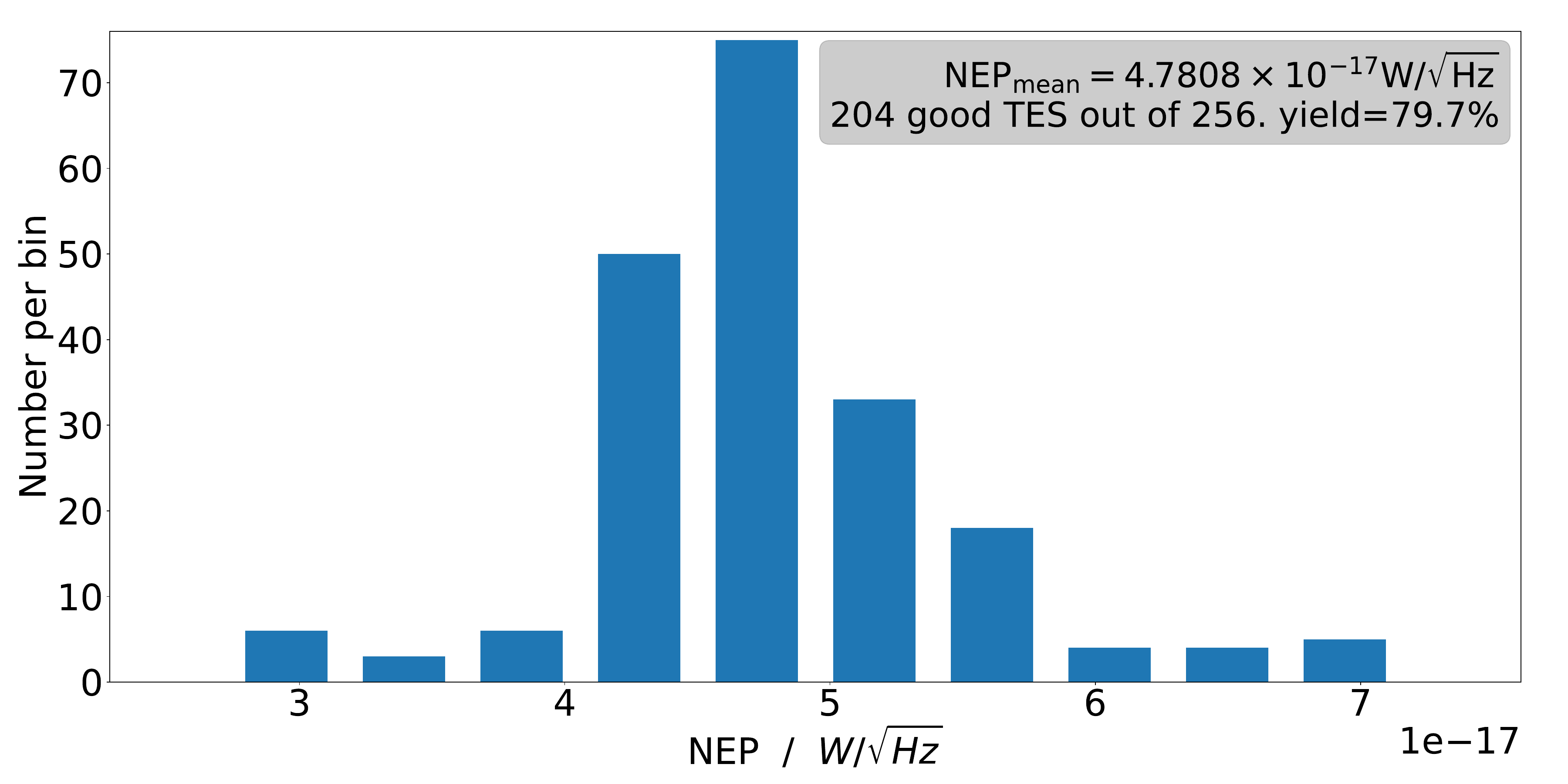}}
 \caption{Histogram of the phonon {noise equivalent power} of
   the full array derived from the fitted parameters. The average phonon NEP is
   $4.8\times10^{-17}~\mathrm{W}/\sqrt{\mathrm{Hz}}$.
   \label{fig:NEPhistogram}}
 \end{figure}

Figure~\ref{fig:NEPhistogram}
shows histograms of the distribution of the
phonon NEP values for the full array derived from the fitted parameters.  There is a strong clustering of
NEP values around $4.8\times10^{-17}~\mathrm{W}/\sqrt{\mathrm{Hz}}$.
The dominance of this value in the histogram is an indication of the homogeneity in the fabrication process of the TES array
(\cite{2019LDT...Marnieros}).

\subsection{Time constants}
The performances of QUBIC have been tested using a monochromatic calibration source~\cite{2020.QUBIC.PAPER3}
To estimate the time constants, the calibration source is modulated in power with a square wave signal with a frequency of 0.6~Hz and a duty cycle of 33\%. The amplitude is chosen to avoid saturation of the detectors while having sufficient signal-to-noise ratio (SNR). The power amplitude on the focal plane is however not constant but corresponds to the synthetic beam. By using a detector located on the calibration source, we checked that the intrinsic rise and fall time is much faster than the expected time constant of the detectors (which is of the order of a few tens of ms). 

%Data were taken on 2020~October~16 with the following configuration:
%\begin{itemize}
%\item    Source modulation 0.6 Hz, square signal, amplitude 0.5 V, offset 1V, Duty Cycle 33\%
%\item    No eccosorb neither on the window nor on the CalSrc horn
%\item    For each of the bias voltages in [1,2,3,4,5] Volts we have taken around 10 minutes data 
%\end{itemize}

\begin{figure}[ht]
 \centering
 \includegraphics[width=0.99\linewidth]{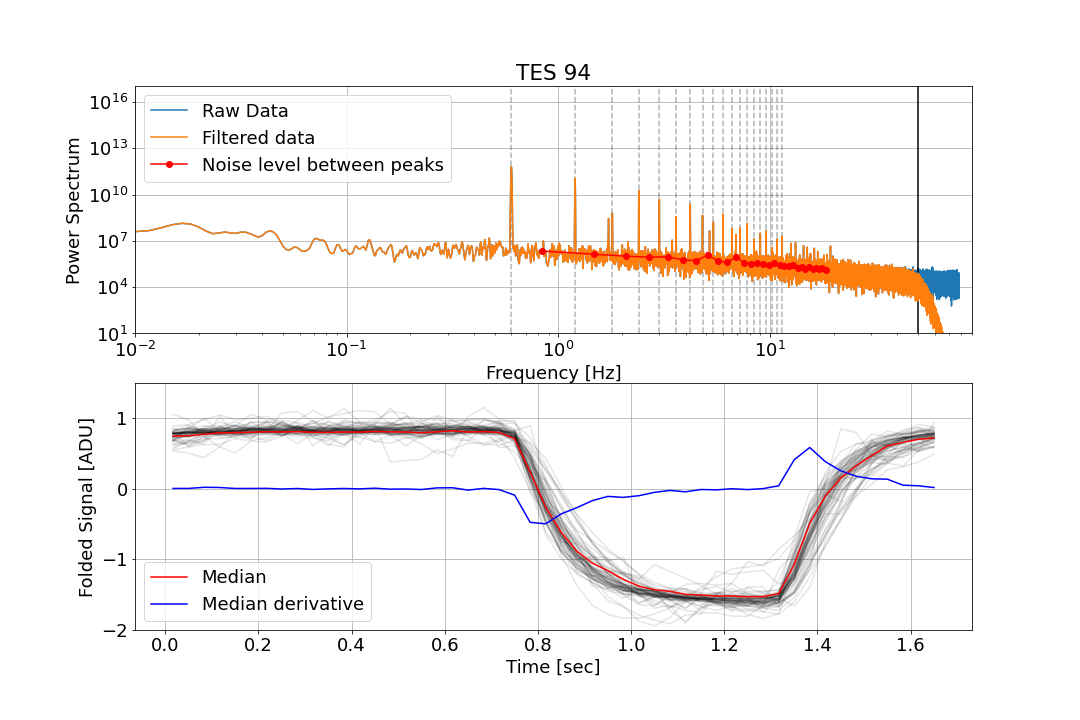}
 \caption{Folded signal for TES 94.  upper: The power spectrum in ADU.  lower: Normalized folded data for some TESs in black, the median of all detectors is shown in red and its derivative in blue.\label{fig:filtering}} \end{figure}

To process the data, we did a very mild low-pass and high-pass filtering as we
do not want the filtering to alter the time constants. We then fold
the data for each TES into one period of the calibration source. The
filtering and the resulting folded signal is shown in Figure~\ref{fig:filtering} for one TES. The signal peaks on the spectrum can be easily seen.

Figure~\ref{fig:filtering} lower shows the normalized (removed average and divided by
RMS) folded data for each TES in black, the median is shown in
red. The derivative is shown in blue and helps finding the first
guess for the start-time of the calibration source shown as a red
dot. Note that no selection has been made at this stage to remove TESs with low SNR.

We then fit each TES folded signal (not normalized -
meaning with its proper amplitude) with a model for the calibration
source signal including a rise time and a fall time. 

%For each TES, we plot the time constant as a function of the bias voltage $V_{TES}$ . 
%not very use full regarding the mean plot 
%\begin{figure}[h!]
%\begin{center}
%\includegraphics[width=0.45\linewidth, keepaspectratio]{figures/TmCst23.png} 
%\includegraphics[width=0.45\linewidth, %keepaspectratio]{figures/TmCst74.png}\\
%\includegraphics[width=0.45\linewidth, keepaspectratio]{figures/TmCst75.png} 
%\includegraphics[width=0.45\linewidth, keepaspectratio]{figures/TmCst94.png}\\
%\caption{{Time constant of four TESs on P87 array in QUBIC.}
%\label{fig:Tcste}}
%\end{center}
%\end{figure}
%An important effect is the diminution of the time constant at lower bias voltage. As $V_{TES}$ decreases, we have an improvement of the time constant on the detector from about 100~ms to 40~ms which is expected from the Electro-Thermal Feedback (ETF) behavior. 

\begin{figure}[ht]
 \centering
 \includegraphics[width=0.80\linewidth]{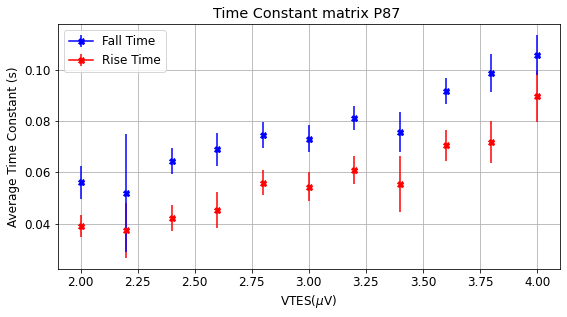}
 \caption{Average value of time constants for rise and fall time as a function of $V_{TES}$.\label{fig:avgtimecst}}
 \end{figure}

Figure~\ref{fig:avgtimecst} shows the average time constants of all TESs as a function of $V_{TES}$. 
The rise time constant appears lower than the fall time indicating again the effect of ETF, but also the fact that we are probably reaching a non-linear regime for most TESs. For small signals, we expect to have a single time constant reaching at most the value of the rise time measured during this sequence, so about 40~ms. This value is enough for QUBIC since the considered scanning speed is about 1~deg/s which leads to a duration of 500~ms for a 30~arcmin beam width.

\subsection{Noise characterizations}
\label{subsect:noise}
Aliasing of the TES Johnson noise is a limitation to Time Domain
Multiplexing performance. Any source of noise before the SQUIDs with a
bandwidth higher than the sampling frequency will be aliased. In Time
Domain Multiplexing, the signal of each detector is averaged during
the duration of measurement $T_{meas}$ which is smaller than the
sampling period $T_s=1/f_s$ by a factor $N_{MUX}$ which is the total
number of pixels readout in the multiplexing scheme. The noise
bandwidth of this averaging is therefore given by
$\Delta f=\frac{1}{2\times T_{meas}}=\frac{f_s\times N_{MUX}}{2}$.
The aliasing leads to an increase of noise given by the square root of the
ratio between the noise bandwidth and the Nyquist frequency $f_s/2$,
that is $\sqrt{N_{MUX}}$.

In QUBIC, the ADC frequency $f_{ADC}=2~MHz$ drives the multiplexing.
The main parameters are therefore: (i) The number of samples $N_s$ to
be read out for each TES, and (ii) the total number of pixels to be
read out.  The maximum number of pixels is equal to $N_{MUX}$ which is
128.  By reducing the number of pixels sampled, the sampling frequency
per pixel is increased.  The sampling frequency per TES is
$f_s=\frac{f_{ADC}}{N_s\times N_{MUX}}$. Typical parameters are
$N_s=100$ and $N_{MUX}=128$ leading to $f_s=156.25~Hz$ and
$\Delta f=10~kHz$.

The SQUID input inductance value is $L_{in}=651~nH$ which leads to
a bandwidth higher than 24~kHz for TES resistance above
100~m$\Omega$. For such resistance values, Johnson noise is increased
by a factor $\sqrt{N_{MUX}}=11.3$.  To overcome this limitation, a
Nyquist inductor can be added in series with the TES. A value of
$L_{Nyq}=15~\mu H$ will reduce the noise bandwidth of Johnson noise to
1~kHz for a 100~m$\Omega$ resistance giving an increase of noise
of~3.6 for the typical parameters. The number of samples $N_s$ can
also be reduced in order to increase the sampling frequency and
further reduce the aliasing. This limitation in 
aliasing was expected for the TD and will be improved for the Full  Instrument by both adding a Nyquist inductor and increasing the sample rate.  The result for the TD is a constraint on NEP of about
$10^{-16}~W/\sqrt{Hz}$, which is a factor 2 higher than the FI requirement, but this sensitivity is acceptable for the QUBIC-TD.

\subsubsection{Noise in normal and superconducting states}
\begin{figure}[ht]
 \centering
{\includegraphics[width=0.49\linewidth]{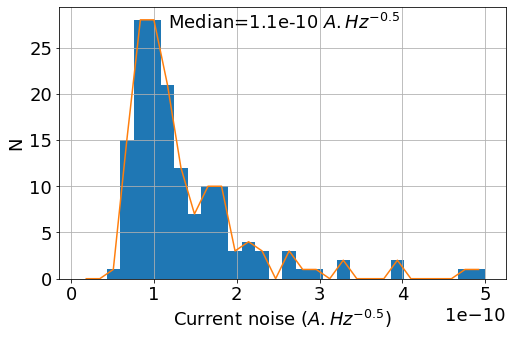}
 \includegraphics[width=0.49\linewidth]{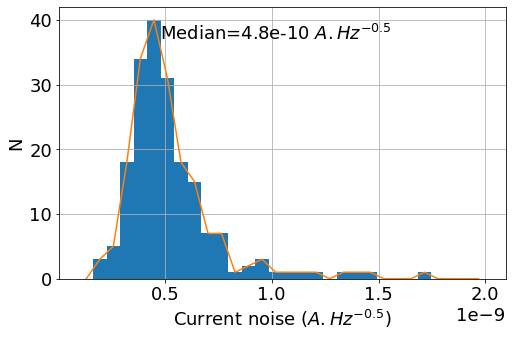}}
 \caption{Histogram of current noise measured between 1~Hz
   and 2~Hz in the normal state (left, 153 total number of TES) and in
   the superconducting state (right, {192} total number of TES).
   \label{InNormalSupra}}
\end{figure}

Figure~\ref{InNormalSupra} shows the histogram of
the measured current noise between 1~Hz and 2~Hz in normal (bias voltage at 8~$\mu$V) and superconducting state of the TES. 
In the normal state, a typical value of 110~$pA/\sqrt{Hz}$ is obtained, compatible with the expectation within a factor of~2 taking into account the aliasing effect.
In the superconducting state, the median current noise is 
470~$pA/\sqrt{Hz}$, compatible with the expectation taking
into account the aliasing effect and the fact that the shunt resistor
and probably part of the parasitic resistance are located on the 1~K
stage, which was cooled to only about 2.6~K during this measurement.

\subsubsection{Noise in the transition}

The detector current noise can be converted into NEP assuming the TES are in strong Electro-Thermal Feedback mode.  In this case, the
TES responsivity $\Re~[A/W]$ is given by the inverse of the TES
voltage, $\Re=\frac{1}{V_{TES}}$. The TES voltage is obtained from the
bias voltage assuming the TES resistance is higher than the shunt
resistance: $V_{TES}=V_{bias}\times 10^{-6}$.

\begin{figure}[ht]
 \centering
{\includegraphics[width=0.49\linewidth]{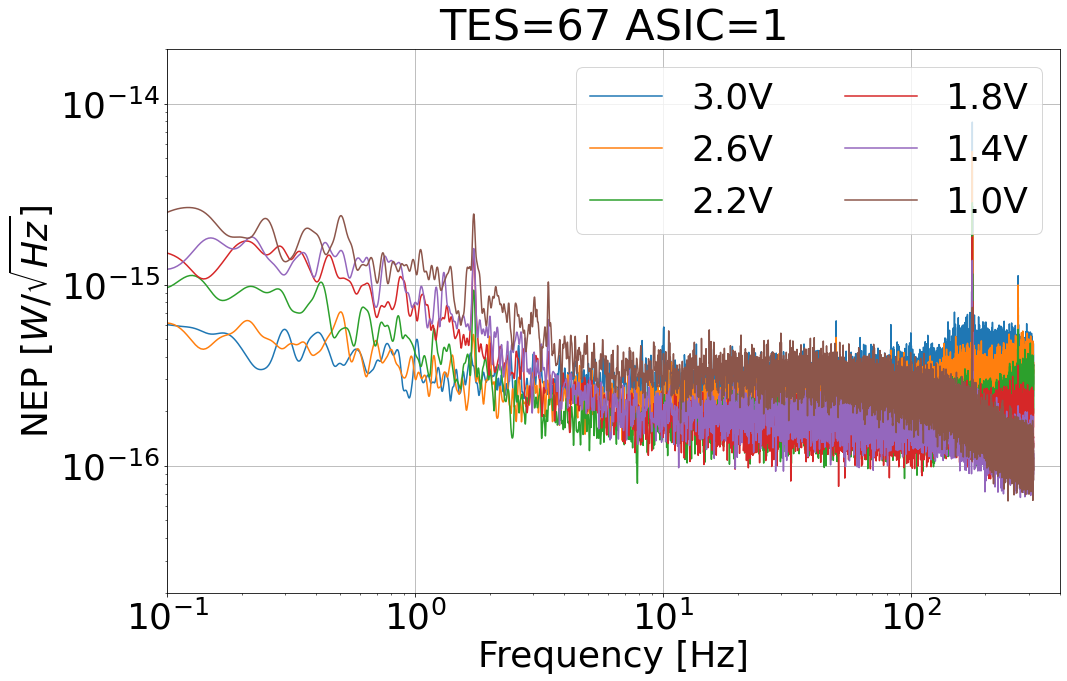}
\includegraphics[width=0.49\linewidth]{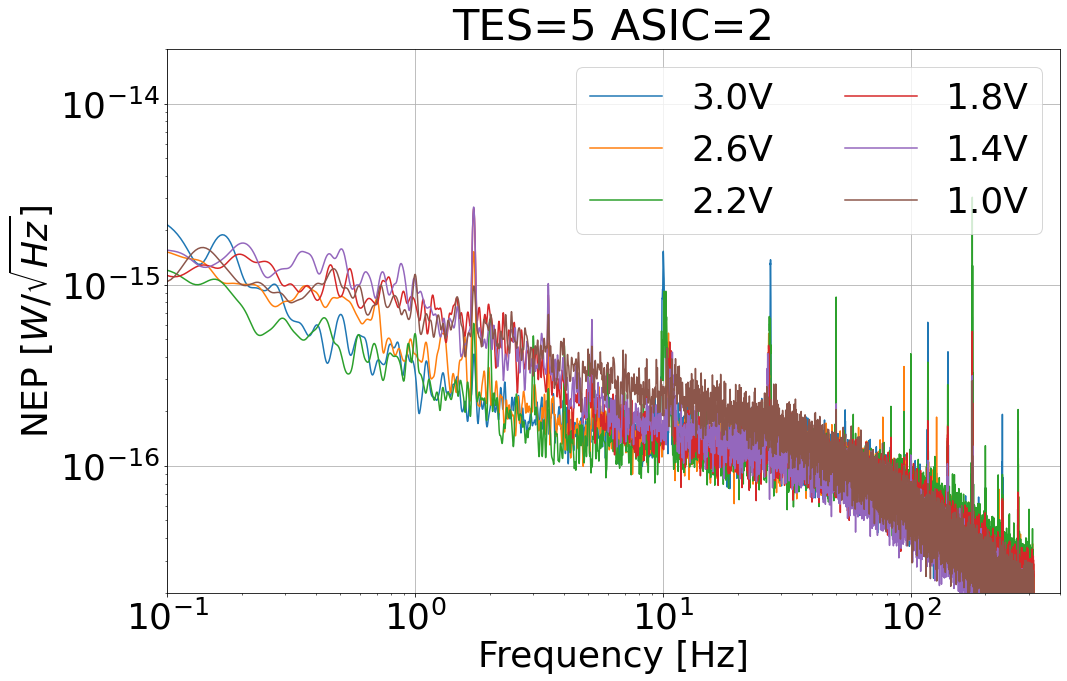}}
 {\includegraphics[width=0.49\linewidth]{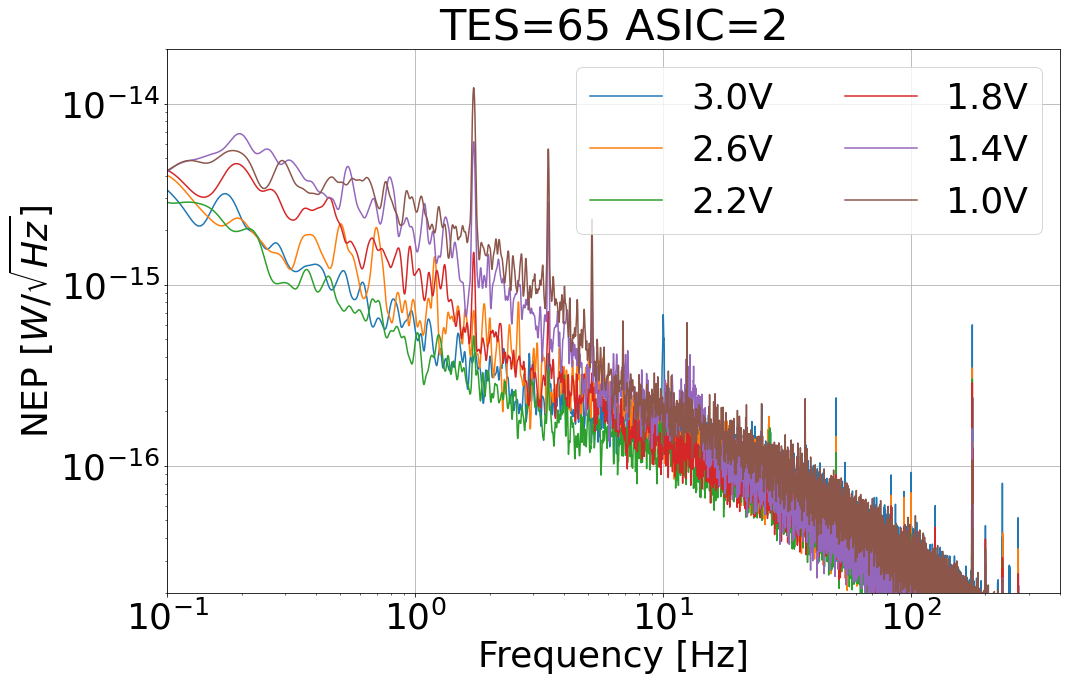}
 \includegraphics[width=0.49\linewidth]{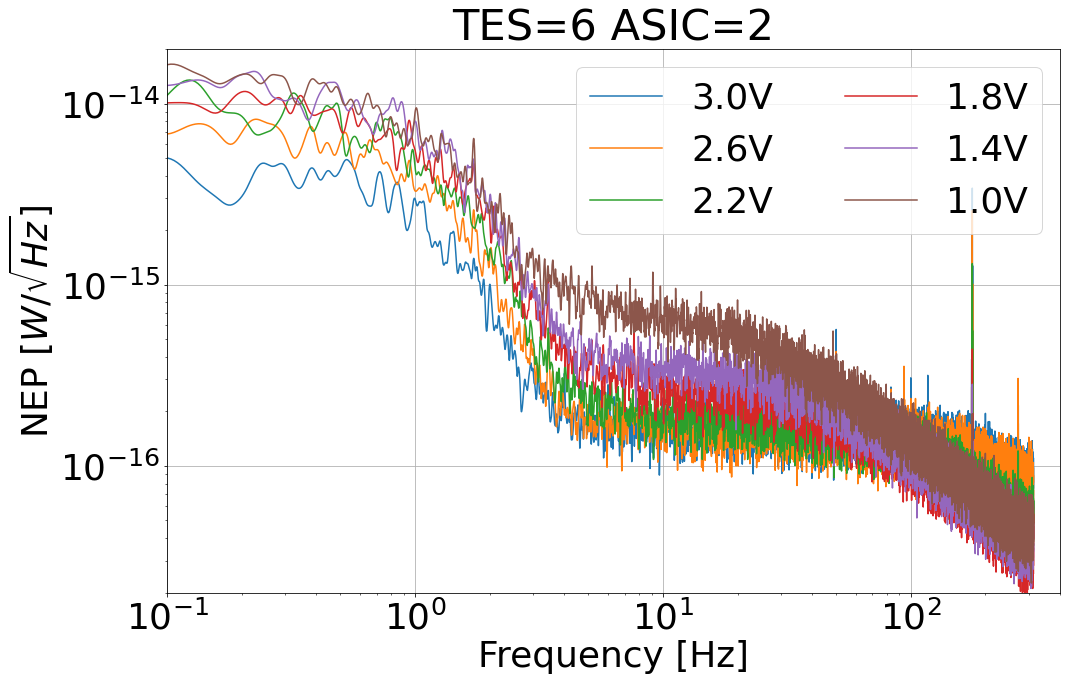}}
 \caption{{NEP spectra} on some channels at
   different bias voltages, from 3~V to 1~V.  This corresponds to the
   ratio of TES and normal resistance ranging from about 60\% to about
   10\%.  
   Note that these measurements were taken at higher frequency sampling by choosing only rows 1 to 8, so $N_{MUX}=32$ which leads to $f_s=625$~Hz.
   \label{spectravbias}}
\end{figure}

Figure~\ref{spectravbias} shows some typical NEP spectra at different bias voltages.  There is clear evidence of a
noise increase at low frequency when decreasing the bias voltage,
which is usually produced by the phonon noise in the TES.  The noise
level is however much higher than expected and it varies between the
TES, as seen in Figure~\ref{spectravbias}. This elevated level has further been attributed to a high sensitivity to microphonics from the pulse tubes (PT) as demonstrated in the following.

A test of sensitivity to pulse tube microphonics was carried out
by stopping the two units for a few minutes.  An example timeline and
associated time-frequency analysis is shown in
Figure~\ref{PTONOFF}. The noise level below few Hz is reduced when both PTs are off while it remains the same at higher frequency. This frequency range where a noise improvement is measured corresponds to the detector
bandwidth.  The induced parasitic signal is therefore thermal on the
detector. The remaining excess of low frequency noise when both PTs are off is attributed to temperature drift.
%Figure~\ref{in_comp_PT_ON_OFF} shows a histogram of the
%measured current noise between 1~Hz and 2~Hz with PTs on and off. The
%distribution appears more spread-out when PTs are on.   The median
%value is close to the noise level measured in the normal state.

\begin{figure}[ht]
 \centering
 {\includegraphics[width=0.9\linewidth]{figures/TES25_57_2020-01-16_105312_new.png}}
 \caption{Examples of timeline in power and corresponding time-frequency analysis (in log of NEP) for two TESs (left: TES 25 and right: TES 57). The two pulse tubes are OFF between $\sim240$~s and
   $\sim420$~s.
   \label{PTONOFF}}
\end{figure}

%\begin{figure}[ht]
% \centering
% \includegraphics[width=0.7\linewidth]{in_comp_PT_ON_OFF}
% \caption{Histogram of current noise measured between 1~Hz and 2~Hz in
%   the transition ($V_{bias}=1.5V$) with PTs ON and OFF.
%   \label{in_comp_PT_ON_OFF}}
%\end{figure}

\begin{figure}[ht]
 \centering
 {\includegraphics[trim=2cm 0cm 2cm 1cm, clip=true, width=0.49\linewidth]{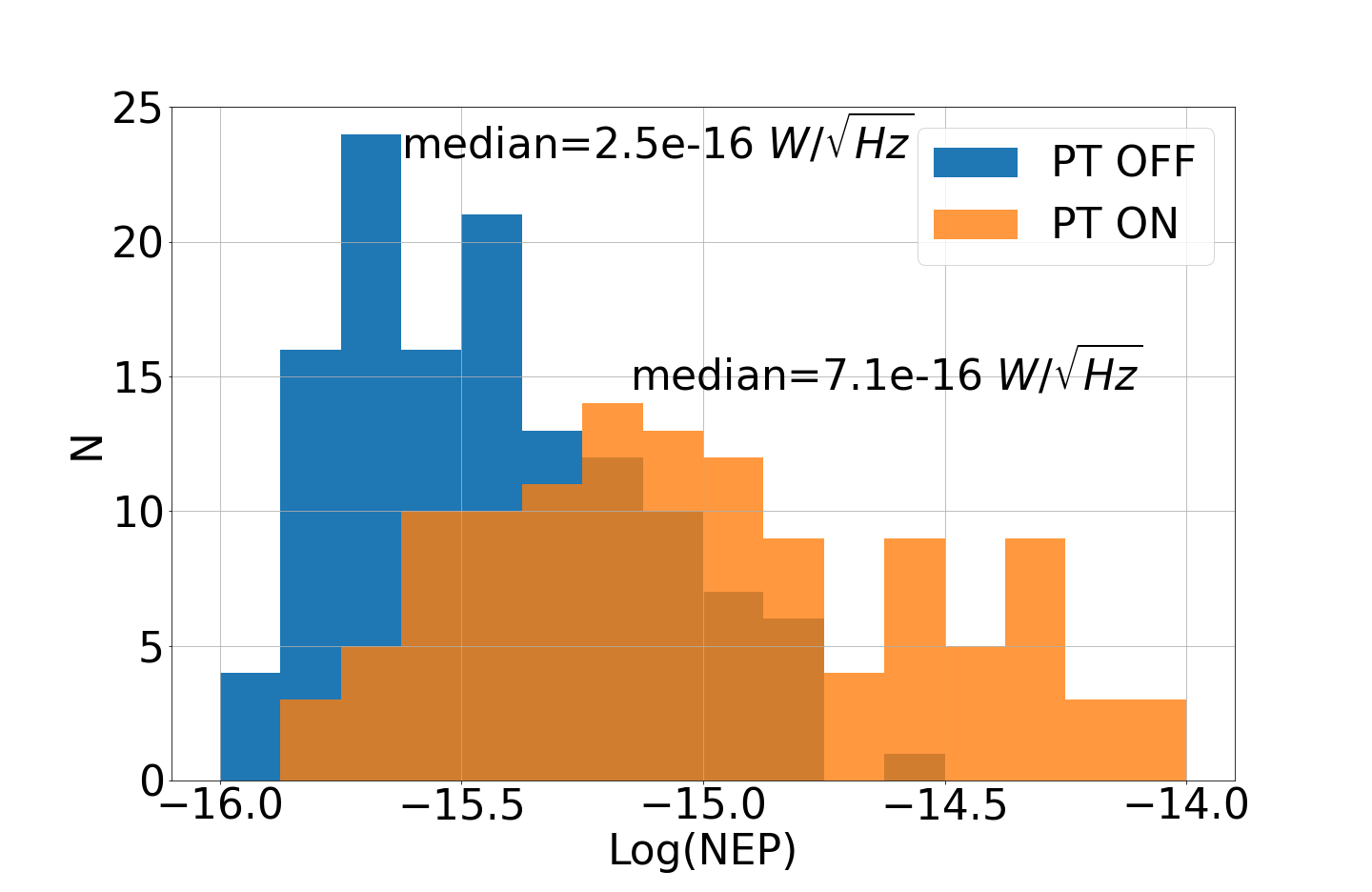}
\includegraphics[trim=2cm 0cm 2cm 1cm, clip=true, width=0.49\linewidth]{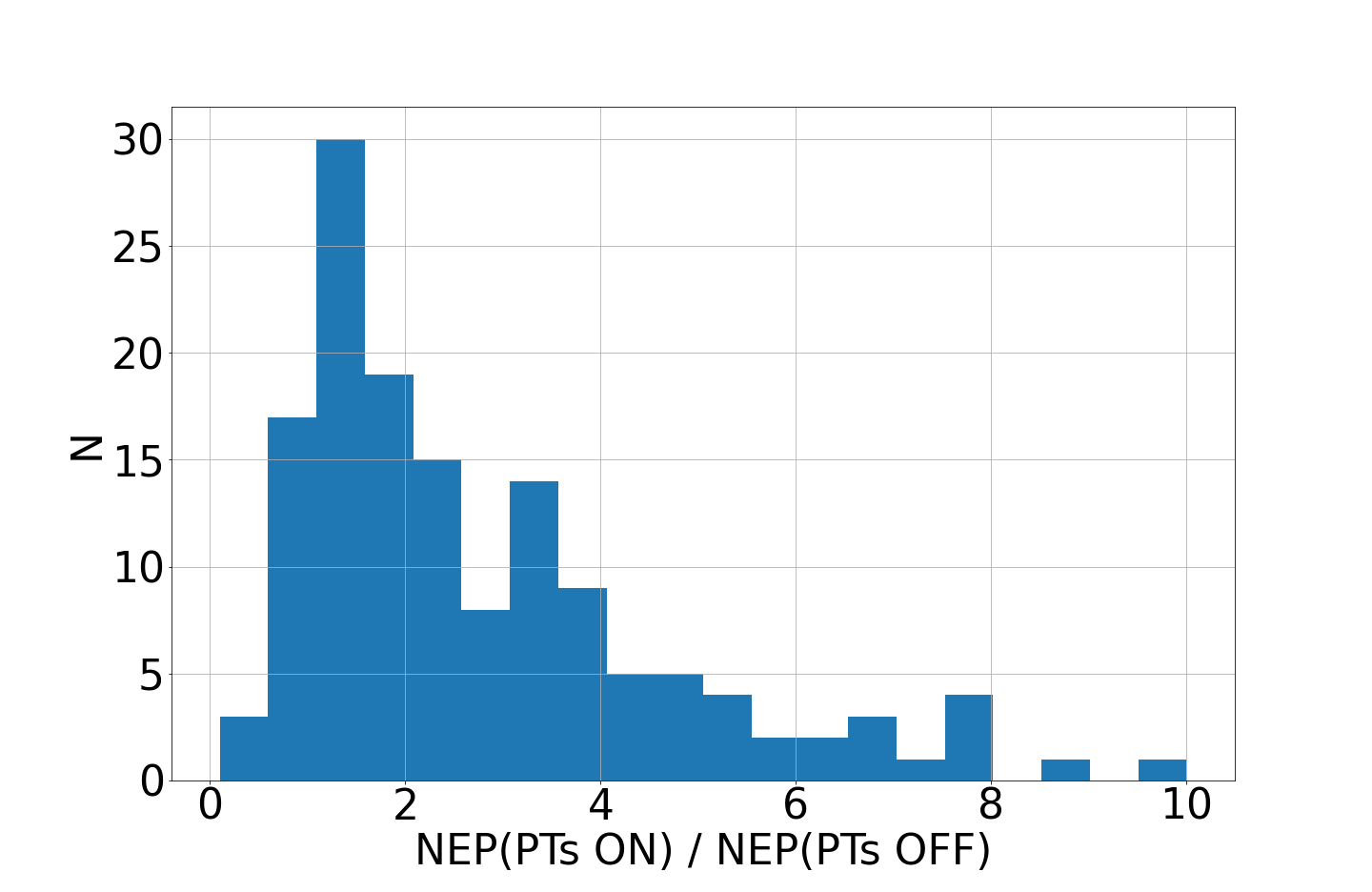}}
 \caption{Left: Histogram of NEP measured between 1~Hz and 2~Hz in the
   transition ($V_{bias}=1.5V$) with PTs ON and OFF. The response is
   assumed to be given by $1/V_{TES}$. The total number of TES are
   130 and 120 respectively. Right: Histogram of the ratio of NEP with
   PTs ON and NEP with PT OFF. The total number of TES is 143.
   \label{NEP_comp_PT_ON_OFF}}
\end{figure}

\begin{figure}[ht]
 \centering
 \includegraphics[width=0.9\linewidth]{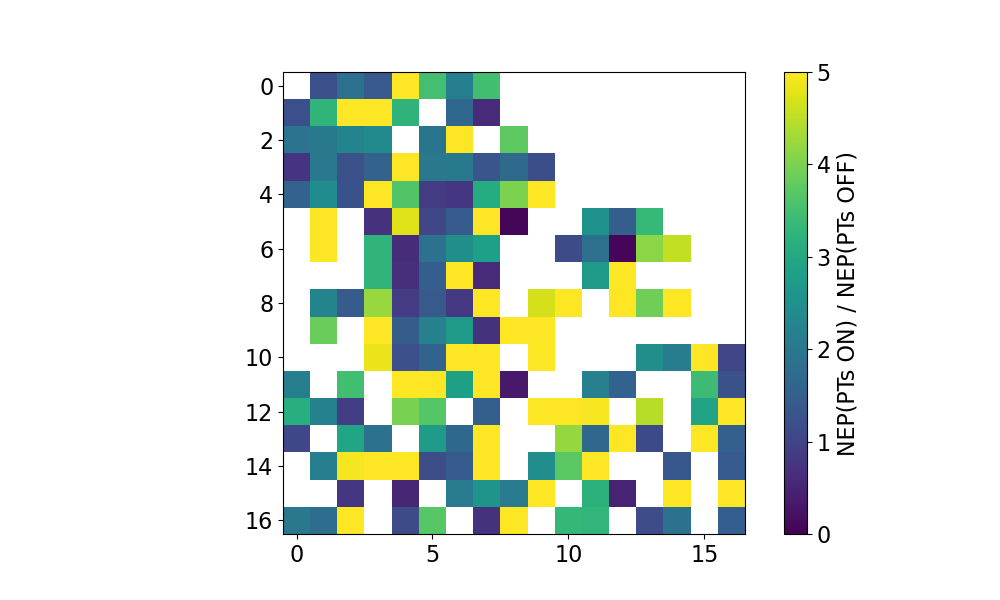}
 \caption{Map of the NEP ratio between PTs on and off. No clear pattern is visible, as one would expect from wafer mechanical resonances.
   \label{map_Rapp_NEPs}}
\end{figure}

Figure~\ref{NEP_comp_PT_ON_OFF} (left) shows the distribution in NEP for
two cases: PTs on or off. It appears that the median NEP when the PT are on is about 3 times higher than when they are off. 
%2 cases are off from specification by a factor 7 and 2.5
%respectively. 
%Some optimizations could be further done in terms of biasing the detectors but
We are clearly dominated by the PT
microphonics. The distribution of the NEP ratio between PTs on and off
is presented in Figure \ref{NEP_comp_PT_ON_OFF} right and Figure
\ref{map_Rapp_NEPs} shows the degradation of noise because of the PTs
on the TES array. If there are mechanical resonances on the wafer, we expect to measure an increase of excess noise in specific locations and most probably in the middle of the array. It is not clear at this stage if we see here some
mechanical specific location on the wafer.

\begin{figure}[ht]
 \centering
    {
    \begin{minipage}[t]{\linewidth}
    \hspace*{0.8cm}\includegraphics[scale =0.3]{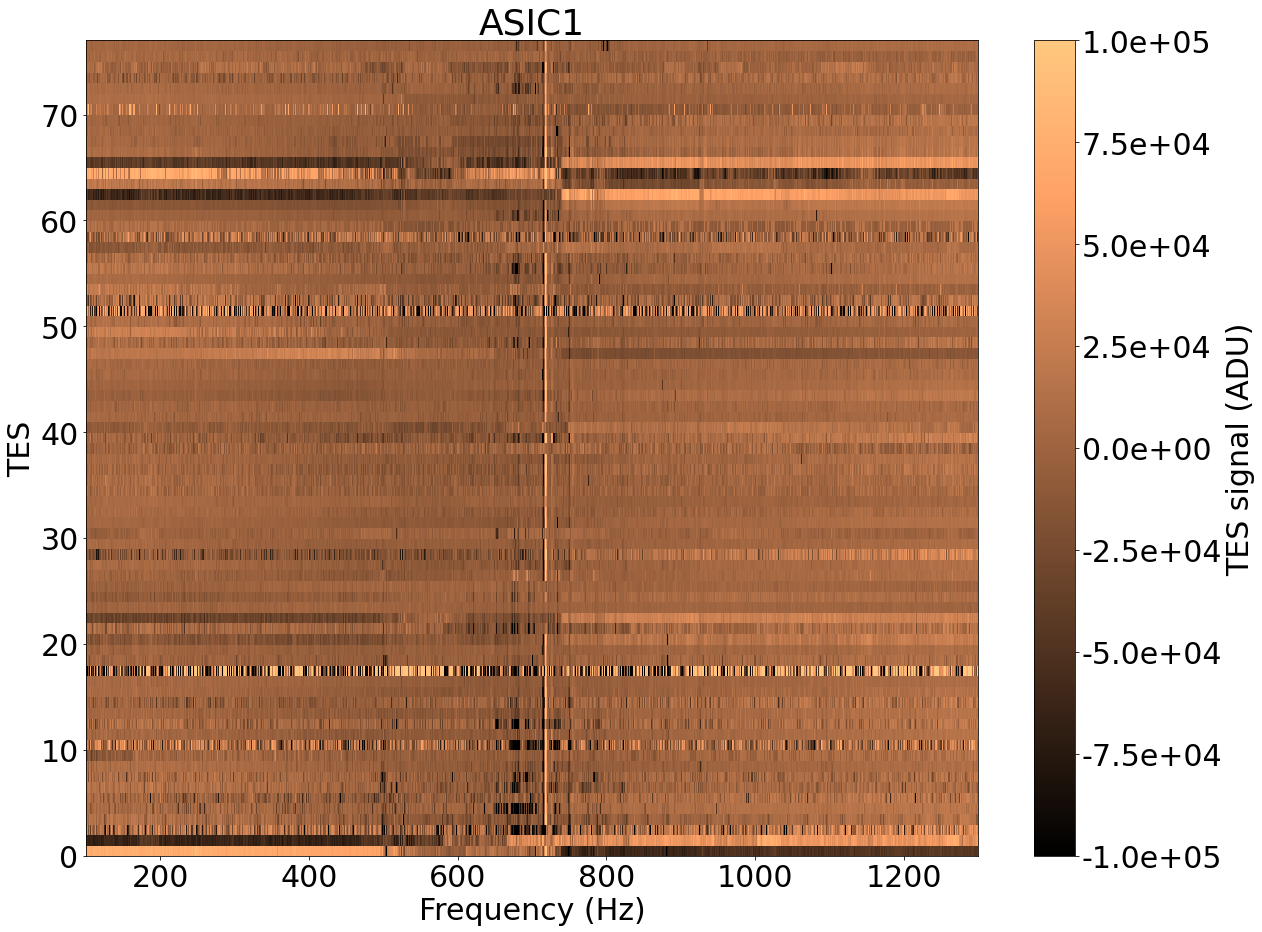}
    \includegraphics[scale =0.24, ]{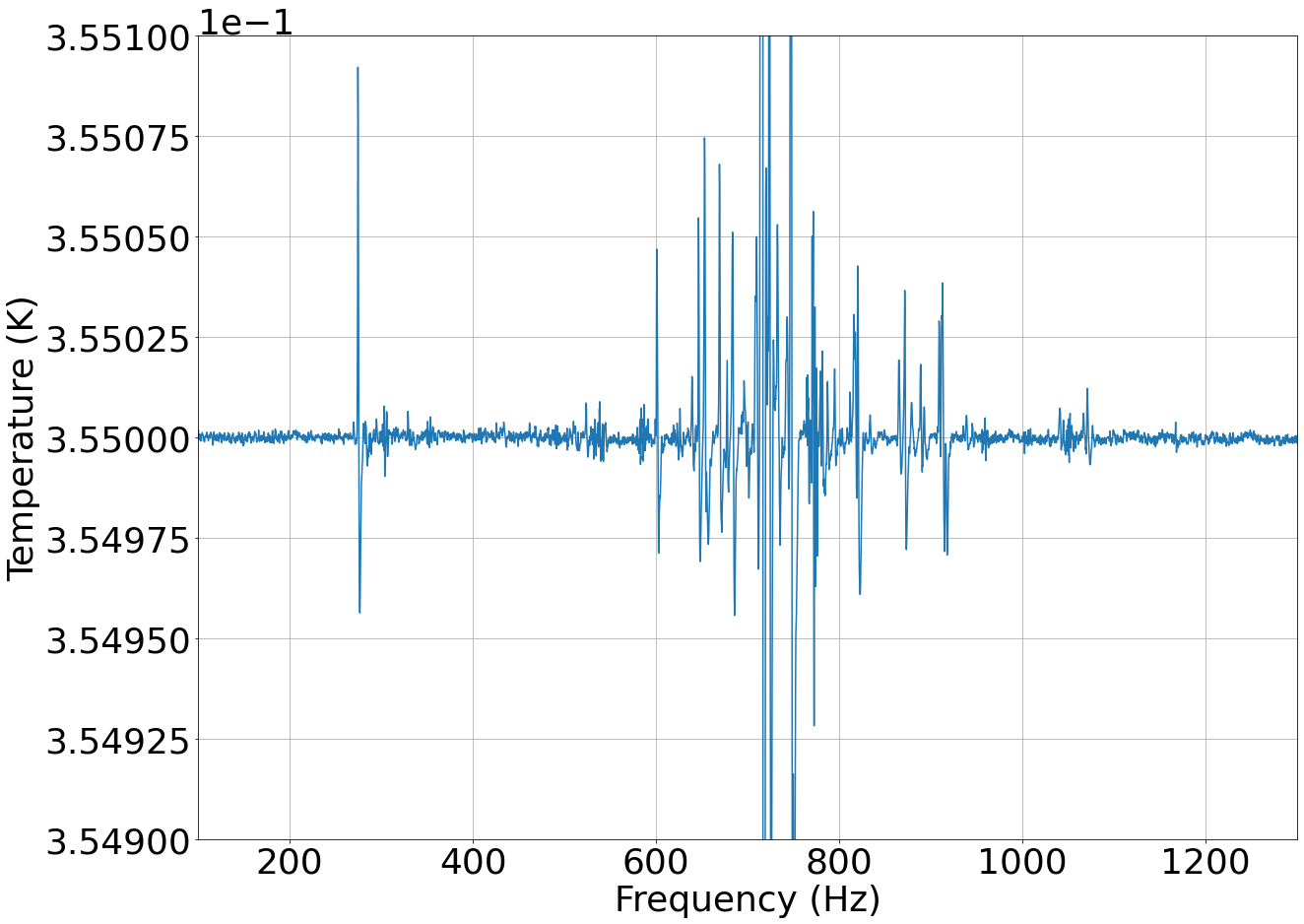}
    \end{minipage}}
  \caption{Top : Time ordered signals in ADU of some TES with the time axis converted in frequency of the mechanical excitation. Bottom: Temperature of the TES
    stage as a function of the frequency of excitation. The graphs
    have been adjusted to share the same x-axis. At mechanical {excitation} frequencies between about 600~Hz and 800~Hz, resonances are clearly seen on the TES signals and in the TES stage temperature.
   \label{fig:NEPtimelines}}
\end{figure}

\begin{figure}[ht]
 \centering
 \includegraphics[width=0.9\linewidth]{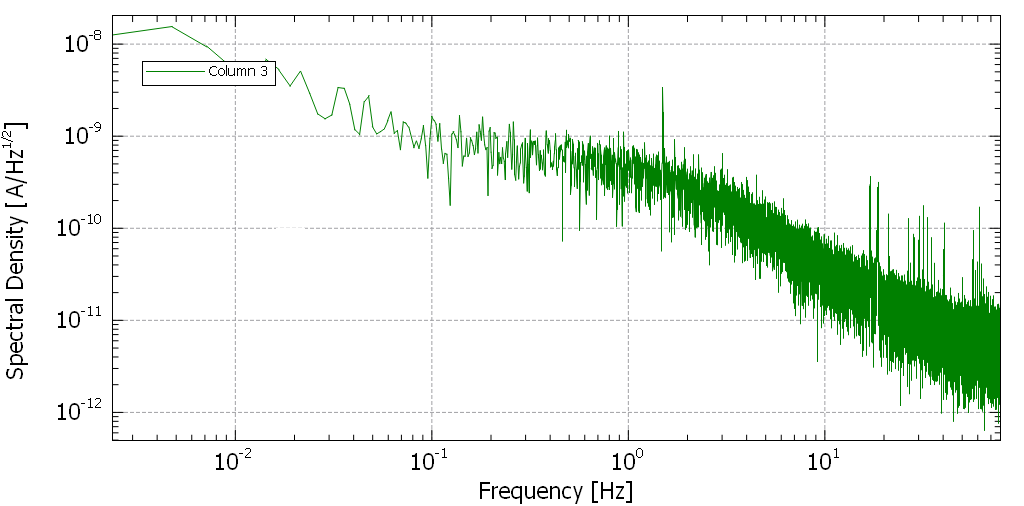}
 \caption{Spectra of TES 96 showing the 1.5~Hz signal from the Amplitude Modulation of the mechanical excitation at 251~Hz. This modulation frequency is not seen off resonance.
   \label{Microphonics_TES96}}
\end{figure}

The origin of these perturbations was investigated.  We checked
from temperature stability measurements that it is not due to thermal
fluctuations of the TES or of the 1~K stage.  The interpretation is the
following: The pulse tube vibrations are exciting mechanical
resonance on the TES support structure but also on the TES
themselves. This mechanical resonance further dissipates heat on
different parts of the system.  This assumption is supported by 3~arguments:
\begin{enumerate}
\item In the timelines of Figure~\ref{PTONOFF} after the PTs are switched off,
  we see a small increase in the TES power which is due to a small
  cooling of the detector, before heating up due to background
  increase.
\item We excited mechanically the cryostat with a speaker connected to
  an audio amplifier and a sine wave generator sweeping from 100~Hz to
  1300~Hz in one hour. Figure~\ref{fig:NEPtimelines} shows signals of TES and of the TES stage thermometer as a function of the excited
  frequency.  Resonances are clearly seen, especially around 700~Hz, probably due to a mechanical resonance.
\item With the same setup, we excited the cryostat at a resonance
  (251~Hz) but the sine wave is modulated in amplitude at 1.5~Hz with
  50\% depth.  Figure~\ref{Microphonics_TES96} shows that this 1.5~Hz
  is seen directly by the TES. When changing the frequency of
  resonance (238~Hz for instance), the 1.5~Hz line disappeared from
  the TES spectra.
\end{enumerate}
We are therefore seeing some heat dissipation produced mainly by the
PT vibrations.  The environment could also contribute to a lesser
extent, for example the traffic on the road nearby.

%\begin{figure}[ht]
% \centering
% \includegraphics[width=0.9\linewidth]{Rapp_NEPs}
% \caption{Histogram of the ratio of NEP with PTs ON and NEP with PT OFF. The NEP were measured between 1Hz and 2Hz in the transition ($V_{bias}=1.5V$). The response is assumed to be given by $1/V_{TES}$. The total number of TESs is 143.
%   \label{Rapp_NEPs}}
%\end{figure}

A better mechanical decoupling of the two PTs is needed to overcome
this problem. The current thermal straps on the 40~K cold heads are made of thin copper plates which are soft in only one direction. Very soft copper braids will
replace these thermal strap to the 40~K shield. The 4~K cold
head is already thermally connected to the 4~K shield with very soft
copper braids.  On the cryostat itself, a soft bellows between the PT
and the structure can be added but this needs a detailed study. It
should be noted that microphonics is a common problem for PT systems
but the effect depends on the detailed mechanical configuration of the
setup. This explains why such a strong effect was not seen at the
sub-system level. This effect is described by
\cite{2008SPIE.7020E..05D,2010SPIE.7741E..1RS,2018JInst..13.8009M}, and~\cite{2019RScI...90e5107G}.

\clearpage
\section{Conclusion}
\label{sec:conclusion}

The QUBIC detection chain based on TES and SQUID, has reached an important milestone. We demonstrated an overall yield of approximately 80$\%$ of working detectors (TESs and SQUIDs included), a thermal decoupling compatible with a phonon noise of about $5\times10^{-17}~\mathrm{W}/\sqrt{\mathrm{Hz}}$ at 410~mK critical temperature, and a time constant of about 40~ms which is enough for the instrument. The QUBIC sensitivity is however currently limited to $2\times10^{-16}~\mathrm{W}/\sqrt{\mathrm{Hz}}$ by microphonic noise and aliasing in the readout electronics. The former will be soon improved by mechanically decoupling the first stages of the pulse tubes. The aliasing of the detector noise will be further improved by increasing the sampling frequency and adding Nyquist inductors to reduce the noise bandwidth of the detectors.

%\textcolor{blue}{
%Some concluding remarks.\\
%SQUIDs:  $\sim90\%$ yield\\
%NEP$_\mathrm{phonon}$: $\sim5\times10^{-17}\mathrm{W}/\sqrt{\mathrm{Hz}}$\\
%NEP: $\sim10^{-16}\mathrm{W}/\sqrt{\mathrm{Hz}}$ for the TD due to limitations in the readout electronics which will be improved in the Full Instrument, especially regarding the SQUID physical temperature and the microphonics.\\
%TES yield $\sim80\%$ including SQUID yield which doesn't necessarily overlap (ie. bad SQUID with bad TES).\\
%microphonics:  dominated by pulse tube microphonics.  working on vibration isolation solutions.\\
%linearity:\\
%magnetic field pickup\\
%improvements for the Full Instrument. confident that we will have the required performance.
%}

\acknowledgments

QUBIC is funded by the following agencies. France: ANR (Agence Nationale de la
Recherche) 2012 and 2014, DIM-ACAV (Domaine d’Intérêt Majeur-Astronomie et Conditions d’Apparition de la Vie), CNRS/IN2P3 (Centre national de la recherche scientifique/Institut national de physique nucléaire et de physique des particules), CNRS/INSU (Centre national de la recherche scientifique/Institut national et al de sciences de l’univers). Italy: CNR/PNRA (Consiglio Nazionale delle Ricerche/Programma Nazionale Ricerche in
Antartide) until 2016, INFN (Istituto Nazionale di Fisica Nucleare) since 2017.  Argentina: MINCyT (Ministerio de Ciencia, Tecnología e Innovación), CNEA (Comisión Nacional de Energía Atómica), CONICET (Consejo Nacional de Investigaciones Científicas y Técnicas).
 
D. Burke and J.D. Murphy acknowledge funding from the Irish Research Council under the Government of Ireland Postgraduate Scholarship Scheme.  D. Gayer and S. Scully acknowledge funding from the National University of Ireland, Maynooth. D. Bennett acknowledges funding from Science Foundation Ireland.

\clearpage
\bibliographystyle{ieeetr}
\bibliography{qubic}  

\end{document}